\documentclass[a4paper,11pt,reqno]{amsart}

\usepackage{geometry}
\usepackage{graphicx}
\usepackage{indentfirst,csquotes}

\usepackage[shortlabels]{enumitem}
\usepackage{graphicx}
\usepackage{amsmath,amsthm,amssymb,mathtools,thmtools}
\numberwithin{equation}{section}

\usepackage{physics2}
\usephysicsmodule{ab}
\usepackage{lscape}
\theoremstyle{plain}

\theoremstyle{definition}

\theoremstyle{remark}

\usepackage{CJKutf8} 

\usepackage{hyperref}
\usepackage[color,final]{showkeys}

\definecolor{refkey}{rgb}{0.1,0.1,0.7}
\definecolor{citekey}{rgb}{0.1,0.1,0.7}
\definecolor{labelkey}{rgb}{0.1,0.1,0.7}

\begin{document}
\renewcommand{\sc}{\scshape}

\title{Bootstrap-Based Estimation and Inference for Measurement Precision under ISO 5725}
\author[J. Takeshita]{Jun-ichi Takeshita${}^{1,2}$}
\author[K. Morita]{Kazuhiro Morita${}^2$}
\author[T. Suzuki]{Tomomichi Suzuki${}^2$}

\address{${}^1$Research Institute of Science for Safety and Sustainability, National Institute of Advanced Industrial Science and Technology (AIST), Tsukuba, Japan}
\email{jun-takeshita@aist.go.jp}
\address{${}^2$Department of Industrial and System Engineering, Faculty of Science and Technology, Tokyo University of Science, Noda, Japan}

\date{\today}
\begin{abstract}
  The ISO 5725 series frames interlaboratory precision through repeatability, between-laboratory, and reproducibility variances, yet practical guidance on deploying bootstrap methods within this one-way random-effects setting remains limited.
  We study resampling strategies tailored to ISO 5725 data and extend a bias-correction idea to obtain simple adjusted point estimators and confidence intervals for the variance components.
  Using extensive simulations that mirror realistic study sizes and variance ratios, we evaluate accuracy, stability, and coverage, and we contrast the resampling-based procedures with ANOVA-based estimators and common approximate intervals.
  The results yield a clear division of labor: adjusted within-laboratory resampling provides accurate and stable point estimation in small-to-moderate designs, whereas a two-stage strategy—resampling laboratories and then resampling within each—paired with bias-corrected and accelerated intervals offers the most reliable (near-nominal or conservative) confidence intervals.
  Performance degrades under extreme designs, such as very small samples or dominant between-laboratory variation, clarifying when additional caution is warranted.
  A case study from an ISO 5725-4 dataset illustrates how the recommended procedures behave in practice and how they compare with ANOVA and approximate methods.
  We conclude with concrete guidance for implementing resampling-based precision analysis in interlaboratory studies: use adjusted within-laboratory resampling for point estimation, and adopt the two-stage strategy with bias-corrected and accelerated intervals for interval estimation.

  \noindent
  \textbf{key words:} ISO 5725; interlaboratory studies; variance components; bootstrap; one-way random effects; repeatability; reproducibility
\end{abstract}

\maketitle

\section{Introduction}
\label{sec:intro}
The ISO 5725 series specifies how interlaboratory studies should be conducted and how their results should be analyzed, and defines the repeatability (within-laboratory), between-laboratory, and reproducibility (the sum of the within-laboratory and between-laboratory variances) variances as key precision measures.
Recent revisions have strengthened computer-based methods for estimating these measures; for example, restricted maximum likelihood (REML) has been incorporated into variance-component estimation.
However, despite the availability of several computer-based approaches, their application to interlaboratory studies remains limited.

Previous studies have examined computer-based variance-component estimation and confidence-interval construction in hierarchical or nested designs.
For example, Wiley~\cite{wiley2001BootstrapStrategiesVariance} investigated methods for estimating variance components in two-way analysis of variance (ANOVA) without replication.
Rostron \textit{et al.}~\cite{rostron2020Confidenceintervalsrobust} proposed robust approaches for constructing confidence intervals in three-stage nested experiments.
In addition, resampling methods have been recognized as promising tools because they do not require distributional assumptions and can accommodate a wide range of measurement data types.

However, several important uncertainties remain unresolved.
Because the structure of ISO 5725-style interlaboratory data is essentially equivalent to a one-way ANOVA with replication, existing computer-based methods for variance-component estimation cannot be directly applied.
Existing methods were developed for experimental structures and objectives fundamentally different from those considered in ISO 5725.
In particular, the practical limitations and comparative performance of resampling schemes have not been systematically evaluated for the repeatability, between-laboratory, and reproducibility variances defined in ISO 5725.
It remains unclear which resampling strategies yield stable point estimators and reliable confidence intervals under realistic constraints such as small sample sizes or large variance ratios.
Consequently, a comprehensive investigation is needed to assess preformance of resampling methods specifically within the ISO 5725 framework.

To address these gaps, this study conducts a systematic comparison of variance-component estimators based on five resampling schemes and evaluates three resampling-based confidence-interval procedures within the ISO 5725 framework.
For point estimation, performance is assessed in terms of bias and standard errors (SEs), whereas coverage probability (CP) is adopted as the primary criterion for evaluating confidence intervals.
We examine the reliability of these approaches across a wide range of sample-size configurations and variance-ratio settings that reflect realistic interlaboratory studies.
Based on these results, we identify resampling strategies that yield stable point estimators and reliable confidence intervals for ISO 5725 precision measures, and we provide practical guidance for implementing computer-based variance-component estimation.

\section{Preliminalies}
\label{sec:prel}

\subsection{One-way random-effects model (quantitative measurements)}
\label{sec:one-way-rand-effects}
In the ISO 5725 series, the basic model for quantitative measured values to estimate the precision of a given measurement method is described by the one-way random-effect model~\cite{iso2025ISO572522025}:
\begin{equation}
	Y_{ij} = \mu  + L_i + E_{ij}, \quad i=1, \ldots, k, \ j=1, \ldots, n,
	\label{eq:1}
\end{equation}
where $Y_{ij}$ is the measured value of trial $j$ in laboratory $i$; $\mu$ is a general mean (expectation); $L_i$ is the laboratory component of variation (under repeatability conditions) in laboratory $i$, whose expectation is assumed to be $0$, and whose variance is the between-laboratory variance $\sigma^2_L$; and $E_{ij}$ is a residual error (under repeatability conditions), whose expectation is also assumed to be $0$, and whose variance is the within-laboratory variance $\sigma^2_{ri}$.
We assume that $L_i$ and $E_{ij}$ are independent, and that the number of replicates is identical in all laboratories, denoted by $n$.
Moreover, the within-laboratory variance $\sigma^2_{ri}$ is assumed to be identical across all laboratories and is denoted by the repeatability variance $\sigma^2_r$.
The reproducibility variance $\sigma^2_R$ is defined as $\sigma^2_R:= \sigma^2_L + \sigma^2_r$.
Noted that the definition of reproducibility variance in the ISO 5725 series differs from that used in Gauge R\&R studies~\cite{burdick2005DesignAnalysisGauge}; the reproducibility variance in Gauge R\&R corresponds to the between-laboratory variance $\sigma_L^2$ in ISO 5725, but the definition in the ISO 5725 series is the first to be defined.

Hereafter, the following notion is used:
\begin{align}
  \mathrm{SSA} &:=  n \sum_{i=1}^k \ab (\bar{Y}_{i} - \bar{\bar{Y}} )^2, \ \mathrm{MSA} = \frac{\mathrm{SSA}}{k-1}
        \label{eq:2}\\
  \mathrm{SSE} &:= \sum_{i=1}^k \sum_{j=1}^n \ab (Y_{ij} - \bar{Y}_i )^2, \ \mathrm{MSE} = \frac{\mathrm{SSE}}{k(n-1)},
        \label{eq:3}
\end{align}
where $\bar{Y}_i = \sum_{j=1}^n Y_{ij} / n$ and $\bar{\bar{Y}} = \sum_{i=1}^k \bar{Y}_i /k$.

Practically, the validation of a given measurement method is based on estimates of $\sigma^2_r$, $\sigma^2_L$, and $\sigma^2_R$.
Unbiased estimators of the three variance components are as follows (see e.g., Searle \textit{et al.}~\cite{searle2006Variancecomponents}):
\begin{align}
  \hat\sigma^2_r = \mathrm{MSE}, \ \hat\sigma^2_L = \frac{\mathrm{MSA} -  \mathrm{MSE}}{n}, \text{ and }
    \hat\sigma^2_R = \hat\sigma^2_r + \hat\sigma^2_L.
  \label{eq:4}
\end{align}
Also, these SEs are given by
\begin{align}
\widehat{\mathrm{SE}}(\hat\sigma^2_r) &= \sqrt{\frac{2 \hat\sigma^4_r}{(k(n-1) +2}},
                                        \label{eq:5} \\\
\widehat{\mathrm{SE}}(\hat\sigma^2_L) &= \sqrt{\frac{2}{n^2} \ab[ \frac{ \ab (n \sigma^2_L + \sigma^2_r)^2}{k+1} + \frac{  \sigma^4_r}{k(n-1)+2}] }, \text{ and }
                                        \label{eq:6}\\
\widehat{\mathrm{SE}}(\hat\sigma^2_R) &= \sqrt{\ab[\widehat{\mathrm{SE}}(\hat\sigma^2_r)]^2 + \ab[\widehat{\mathrm{SE}}(\hat\sigma^2_L)]^2 - \frac{2 \sigma^4_r}{kn(n-1) + 2}}.
                                        \label{eq:7}
\end{align}

In particular, it is important to know whether there is no difference between laboratories, that is, whether $\sigma^2_L = 0$ or not.
However, since the estimator of $\sigma^2_L$ is given by the difference between the estimators of $\sigma^2_R$ and $\sigma^2_r$ (Eq.~(\ref{eq:4})), it is difficult to derive a CI for the estimator of $\sigma^2_L$ theoretically.
These shortcomings motivate the bootstrap and bias-correction strategies developed in Sections 3–5.

\subsection{Bootstrap framework}
\label{sec:bootstrp-fram}

Let $Y=(Y_{1},\dots ,Y_{n})$ be a random sample from an unknown distribution $F$.
By the Glivenko–Cantelli theorem, it is well-kwon that the empirical distribution $F_{n}$, which assigns probability $1/n$ to each observation, is a consistent estimator of $F$ (see e.g., van der Vaart~\cite[Chapter 19]{vaart2007Asymptoticstatistics}).
For a parameter $\theta=\theta(F)$ with an estimator $\hat{\theta}=\theta(Y_1, \ldots, Y_n) = \theta(F_{n})$ from the samples, the bootstrap approximates the sampling distribution of $\hat{\theta}$ by repeated resampling from $F_{n}$:
\begin{enumerate}
	\item[(i)] Draw $M$ independent bootstrap samples $y_{m}^{*}=(y_{1m}^{*},\dots ,y_{nm}^{*})$ ($m=1,\dots ,M$) by sampling with replacement from $y$, a real value of $Y$.
	\item[(ii)] For each sample, compute the replicate statistic $\hat{\theta}^{*}_{m}=\theta(y_{m}^{*})$,
        \end{enumerate}
        where and hereafter a superscript ${}^*$ denotes quantities computed from bootstrap resamples.
Then, the bootstrap estimates of the mean and variance of $\hat{\theta}$ are, respectively, as follows:
\begin{align}
	\bar{\theta}^{*} =\frac{1}{M}\sum_{m=1}^{M}\hat{\theta}^{*}_{m} \quad \text{and} \quad 
	\widehat{\operatorname{Var}}_{\mathrm{BS}}(\hat{\theta}) = \frac{1}{M-1}\sum_{m=1}^{M}\left(\hat{\theta}^{*}_{m}-\bar{\theta}^{*}\right)^{2}.
	\label{eq:8}
\end{align}
Because the bias of $\hat\theta$ is estimated by $\widehat{\mathrm{bias}}_{\mathrm{BS}}=\bar\theta^*-\hat\theta$, the bias-corrected estimator and the bootstrap SE are, respectively,  yielded by
\begin{align}
  \hat\theta_{\mathrm{cor}}=2\hat\theta-\bar\theta^* \quad \text{and} \quad 
 \widehat{\mathrm{SE}}_{\mathrm{BS}}=\sqrt{\widehat{\mathrm{Var}}_{\mathrm{BS}}}.
   \label{eq:9}
\end{align}

\section{Bootstrap strategies and bias-corrected estimators}
\label{sec:bootstr-strat-bias-c}

\subsection{Five resampling schemes}
\label{sec:resampl-schem}

This study deals with five types of resampling methods.
\begin{enumerate}[leftmargin=*]
\item \emph{Laboratory-level resampling (boot-$i$)}

  This approach involves resampling laboratories.
  For the measured value $y_{ij}$, each $i$ is resampled with replacement while keeping $j$ fixed.
  In this method, only the laboratories are resampled; the measurement results within each laboratory remain unchanged.
  Following the symbols of Wiley~\cite{wiley2001BootstrapStrategiesVariance}, this method is referred to as \emph{boot-$i$} because the subscript $i$ is the subject of resampling.

\item \emph{Single within-laboratory resampling (boot-$j_s$, where the subfix $s$ indicates ``single'')}

  In this method, for $y_{ij}$, each $i$ is fixed and the replicate index $j$ is resampled once with replacement within each laboratory. 
This scheme corresponds to \emph{boot-$j$} described by Wiley~\cite{wiley2001BootstrapStrategiesVariance}; however, we denote it as \emph{boot-$j_s$} in this study to distinguish it from the next resampling scheme.

  \item  \emph{Repeated within-laboratory resampling (boot-$j_r$, where the subfix $r$ indicates ``repeated'')}

    In this approach, for $y_{ij}$, each $i$ is fixed and each $j$ is resampled with replacement for each time.
    Resampling is restricted to within each laboratory, rather than across all measurements in the original sample, to maintain the statistical meaning of each variance component estimate.
    This study uses the abbreviation \emph{boot-$j_r$} to denote this method.

\item \emph{Two-stage hierarchical resampling (boot-$ij_r$) }

  This approach consists of two steps: (1) resampling laboratories with replacement, and (2) resampling measurement results with replacement within each selected laboratory.
  This method is identical to performing \emph{boot-$i$} followed by \emph{boot-$j_r$}.
  This study uses the abbreviation \emph{boot-$ij_r$} to indicate this method.

\item \emph{Double resampling (boot-$ij_s$)}
  
  This method is an approach in which, for $y_{ij}$, both $i$ and $j$ are resampled once with replacement.
This scheme corresponds to \emph{boot-$ij$} described by Wiley~\cite{wiley2001BootstrapStrategiesVariance}; however, we denote it as \emph{boot-$ij_s$} in this study to distinguish it from the previous resampling scheme.
\end{enumerate}

\noindent




\subsection{Applying the Wiley bias-correction formula}
\label{sec:appl-wiley-bias-cor}

Wiley~\cite{wiley2001BootstrapStrategiesVariance} proposed a theoretical correction for bootstrap point estimators under a  two-way random effects model without replication, whose basic model can be written as
\begin{equation}
	Y_{ij} = \mu + A_{i} + B_{j} + E'_{ij}, \quad i = 1, \ldots, k, \ j = 1, \ldots, n.
	\label{eq:10}
\end{equation}
Here, $\mu$ is the overall mean, $A_i \sim N (0, \sigma^2_A)$ and $B_j \sim N (0, \sigma^2_B)$ represent the random main effects, and $E'_{ij} \sim N(0, \sigma^2_{AB})$ is the residual error.
All components are assumed to be independent and identically distributed.

This model can be included the model (\ref{eq:1}).
Indeed, $A_i$, $B_j$, and $E'_{ij}$ should be set $L_i$, $0$, and $E_{ij}$, respectively.
By applying the results of Wiley~\cite[Figures 1--3]{wiley2001BootstrapStrategiesVariance}, we define adjustment formulas for \emph{boot-$i$}, \emph{boot-$j_s$}, and \emph{boot-$ij_r$}, respectively, as follows:
\begin{align}
	\hat\sigma^2_{r:ad} & = \frac{k}{k-1} \bar\sigma^{2*}_{r}, \quad \hat\sigma^2_{L:ad} = \frac{k}{k-1} \bar\sigma^{2*}_{L};
	\label{eq:11}                                                                                                                                                                                                                 \\
	\hat\sigma^2_{r:ad} & = \frac{n}{n-1} \bar\sigma^{2*}_{r}, \quad \hat\sigma^2_{L:ad} := \hat\sigma^2_{L} = \bar\sigma^{2*}_{L} - \frac{1}{n-1} \bar\sigma^{2*}_{r};
	\label{eq:12}                                                                                                                                                                                                                 \\
	\hat\sigma^2_{r:ad} & = \frac{k}{k-1}\cdot \frac{n}{n-1}  \bar\sigma^{2*}_{r}, \quad \hat\sigma^2_{L:ad} := \frac{k}{k-1}\left( \bar\sigma^{2*}_{L} - \frac{1}{n-1} \bar\sigma^{2*}_{r} \right).
	\label{eq:13}
\end{align}
Here, $\bar\sigma^{2*}_{r}$ and  $\bar\sigma^{2*}_{L}$ stand for the bootstrap mean defined in (\ref{eq:8}) with $\theta$ set to $\sigma^2_r$ and $\sigma^2_L$, respectively.
Also, from the similarities of the methods, we use the adjustment formulas of \emph{boot-$j_s$} (\ref{eq:12}) and \emph{boot-$ij_r$} (\ref{eq:13}) for \emph{boot-$j_r$} and \emph{boot-$ij_s$}, respectively.
These adjustment formulas will be verified in the simulation study.


\section{Confidence interval of variance components}
\label{sec:conf-interv-vari-com}

\subsection{Approximate  confidence intervals}
\label{sec:approxi-conf-interv}

Since $e_{ij} \sim N(0,\sigma_r^{2})$, we have
\begin{equation}
	\frac{\mathrm{SSE}}{\sigma_r^{2}}
	= \sum_{i=1}^{k} \sum_{j=1}^{n}
	\frac{\bigl(Y_{ij}-\bar{Y}_{i}\bigr)^{2}}{\sigma_r^{2}}
	\sim \chi^{2}(\phi_E).
	\label{eq:14}
\end{equation}
Therefore, a $100(1-\alpha)\%$ confidence interval for $\sigma_r^{2}$ is
\begin{equation}
	\left[
		\frac{\mathrm{SSE}}{\chi^{2}_{\phi_E,\,\alpha/2}},\;
		\frac{\mathrm{SSE}}{\chi^{2}_{\phi_E,\,1-\alpha/2}}
		\right],
	\label{eq:15}
\end{equation}
where $\chi^{2}_{\nu,p}$ denotes the $p$-th quantile of the chi-square distribution with $\nu$ degrees of freedom.

Confidence intervals of the between-laboratory ($\sigma_L^{2}$) and reproducibility ($\sigma_R^{2}$) variances cannot be obtained in explicit form, but several approximations are known.
Regarding an approximation for between-laboratory variance, Moriguchi's formula~\cite{moriguti1954Confidencelimitsvariance}, which is shown by Boardman~\cite{boardman1974ConfidenceIntervalsVariance} to fit best among several candidates, gives the $100(1-\alpha)\%$ confidence limits
\begin{align}
	\sigma_{L:\mathrm{M,\,lower}}^{2}
	 & =\frac{\mathrm{MSA}}{n}\!
	\left[
		\frac{1}{F_{L}}
		-\frac{\mathrm{MSE}}{\mathrm{MSA}}
		-b_{L}\!
		\left(\frac{\mathrm{MSE}}{\mathrm{MSA}}\right)^{2}
		\right],
	\label{eq:16}                \\
	\sigma_{L:\mathrm{M,\,upper}}^{2}
	 & =\frac{\mathrm{MSA}}{n}\!
	\left[
		\frac{1}{F_{U}}
		-\frac{\mathrm{MSE}}{\mathrm{MSA}}
		+b_{U}\!
		\left(\frac{\mathrm{MSE}}{\mathrm{MSA}}\right)^{2}
		\right],
	\label{eq:17}
\end{align}
with
\begin{align}
	b_{L} & =\frac{F_{L}}{\phi_{E}}
	\left(\frac{\phi_{A}F_{L}}{2}-\frac{\phi_{A}-2}{2}\right),
	\label{eq:18}                   \\
	b_{U} & =\frac{F_{U}}{\phi_{E}}
	\left(\frac{\phi_{A}-2}{2}-\frac{\phi_{A}F_{U}}{2}\right),
	\label{eq:19}
\end{align}
and $F_{L}=F(\phi_{A},\infty,\alpha/2)$, $F_{U}=F(\phi_{A},\infty,1-\alpha/2)$ denoting the upper $\alpha/2$ and $(1-\alpha/2)$ quantiles of the $F$-distribution with $\phi_{A}$ and $\infty$ degrees of freedom.

For the reproducibility variance,
\begin{equation}
	\hat\sigma_{R}^{2}
	=\hat\sigma_{r}^{2}+\hat\sigma_{L}^{2}
	=\frac{\mathrm{MSA}}{n}+\Bigl(1-\frac{1}{n}\Bigr)\mathrm{MSE},
	\label{eq:20}
\end{equation}
a Satterthwaite approximation yields the interval
\begin{equation}
	\left[
		\frac{\phi^{\ast}\hat\sigma_{R}^{2}}{\chi^{2}(\phi^{\ast},\,\alpha/2)},\;
		\frac{\phi^{\ast}\hat\sigma_{R}^{2}}{\chi^{2}(\phi^{\ast},\,1-\alpha/2)}
		\right],
	\label{eq:21}
\end{equation}
where the effective degrees of freedom are
\begin{equation}
	\phi^{\ast}
	=\frac{\bigl[\mathrm{MSA}+(n-1)\mathrm{MSE}\bigr]^{2}}
	{\mathrm{MSA}^{2}/\phi_{A}+(n-1)^{2}\mathrm{MSE}^{2}/\phi_{E}}.
	\label{eq:22}
\end{equation}

\subsection{Resampling-based confidence intervals}
\label{sec:resaml-based-conf-in}

The paper applies three bootstrap procedures used to construct two-sided $100(1-\alpha)\%$ confidence intervals (CIs) for a parameter $\theta$.

\noindent
(a) \emph{Standard-normal (Wald-type) interval}

Ivolking the central-limit theorem, $(\hat\theta - \theta) \mathrm{SE}(\hat\theta) \sim N(0,1)$ for large $n$.
Replacing the unknown stadard error by $\widehat{\mathrm{SE}}_{\mathrm{BS}}$ yields
\begin{equation}
	\ab [\hat\theta - z_{1-\alpha/2}\widehat{\mathrm{SE}}_{\mathrm{BS}}, \hat\theta - z_{\alpha/2}\widehat{\mathrm{SE}}_{\mathrm{BS}}],
	\label{eq:23}
\end{equation}
where $z_q$ is the $q$th standard-normal quantile.

\smallskip

\noindent
(b) \emph{Percentile confidence interval method} (see, e.g., Efron and Tibshirani~\cite[Chapter 13]{efron1994IntroductionBootstrap})

Let $\hat\theta^*_q$ denote the $q$-quantile of a distribution obtained by ordering the $M$ bootstrap estimates $\hat\theta^*_{m} \ (m=1, \ldots, M)$ in ascending order.
Then, the two-sided $100(1-\alpha)\%$ CIs given by the percentile method are given by
\begin{equation}
	\ab [\hat\theta^*_{(\alpha/2)}, \hat\theta^*_{(1-\alpha/2)}].
	\label{eq:24}
\end{equation}

\smallskip

\noindent
(c) \emph{Bias-corrected and accelerated percentile confidence interval method (BCa method)} (see e.g., Efron and Tibshirani~\cite[Chapter 14]{efron1994IntroductionBootstrap}))

The upper and lower limits of $100(1-\alpha)\%$ CIs based on the BCa method are given by
\begin{align}
	\beta_{[\alpha/2]}    := \Phi \ab (z_0 + \frac{z_0 + z_{\alpha/2}}{1 - a (z_0 + z_{\alpha/2})}) \text{ and } 
	\beta_{[1-\alpha/2]}  := \Phi \ab (z_0 + \frac{z_0 + z_{1-\alpha/2}}{1 - a (z_0 + z_{1-\alpha/2})}).
	\label{eq:25}
\end{align}
Here, $z_0$ and $a$ are the bias-correction and acceleration parameters, respectively, which adjust the location and the width and shape of the CI.
These parameters can be estimated as follows:
\begin{align}
	\hat{z}_0  = \Phi^{-1} \ab (\frac{\# \ab \{\hat\theta^*_m \leq \hat\theta\}}{M}), \quad 
	\hat{a}    = \frac{\sum_{m=1}^M \ab (\hat\theta^*_m - \bar\theta^*)^3}{6 \ab [\sum_{m=1}^M \ab  (\hat\theta^*_j - \bar\theta^*)^2]^{3/2}} 
	\ \ab(\bar\theta^* := \frac{1}{M} \sum_{m=1}^M \hat\theta^*_m).
	\label{eq:26}
\end{align}

\section{Simulation study}
\label{sec:simul-study}

\subsection{Environmental settings}
\label{sec:envir-sett}

In order to validate which resampling schemes and confidence intervals are suitable to estimate the precision measures, the procedure for the Monte Carlo simulation is used in this study as described below.

First, we assume a normal distribution for the population with a general mean of $\mu = 0$.
For the variance component parameters, we fixed the within-laboratory variance $\sigma^2_r$ at $1.00$ and established four scenarios with different ratios of between-laborataory variance to within-laboratory variance ($\sigma^2_L / \sigma^2_r$) of $0.25$, $0.50$, $1.00$, and $2.00$. 
The numbers of laboratories $k$ and replicates $n$ for each sample $Y_m$ consisted of $16$ combinations where $k,n \in \{3, 5, 10, 50\}$.
The number of bootstrap iteratios was set to $M=1000$, and that of Monte Carlo replications was set to $1000$.
The confidence level was set to $95\%$ (significance level $\alpha=0.05$).
Therefore, the simulation was conducted for a total of $64$ conditions ($4$ variance ratios $\times$ $16$ sample size combinations).
As a non-resampling benchmark, we also computed the ANOVA estimators (Eqs.~(\ref{eq:4})--~(\ref{eq:7})) and the approximate CIs (Section~\ref{sec:approxi-conf-interv}).

\subsection{Results of the simulation study}
\label{sec:results-simul-study}

The results of this simulation study are presented in the following order: point estimation, SE, and CIs.
In brief, the adjusted \emph{boot-$j_r$} and \emph{boot-$j_s$} provide the most accurate and stable point estimates, while the adjusted \emph{boot-$ij_r$} combined with BCa delivers the most reliable (near-nominal or conservative) CIs.

\subsubsection{Point estimation}
\label{sec:point-estim}

First, we evaluated the bias-corrected estimators defined in (\ref{eq:9}) for the case $\sigma^2_L / \sigma^2_r = 0.50$.
For small samples (e.g., $(k, n) = (5, 5)$), \emph{boot-$j_r$}, \emph{boot-$j_s$}, \emph{boot-$ij_r$}, and \emph{boot-$ij_s$} tended to overestimate $\sigma^2_L$, whereas the \emph{boot-$i$} yielded values closer to the true $\sigma^2_r$.
For the between-laboratory variance, all methods showed a tendency toward underestimation.
This bias diminished more rapidly as the number of replicates $n$ increased than as number of laboratories $k$ increased.

 In particular, with $(k,n)=(5,5)$ and $\sigma_L^2/\sigma_r^2=0.5$, the adjusted \emph{boot‑$j_r$} and \emph{boot-$j_s$} provided $\hat\sigma_{r:\mathrm{ad}}^2\approx 1.01$ and $\hat\sigma_{L:\mathrm{ad}}^2v\approx 0.50$, essentially matching the true values (Table~\ref{tab:PE_k=5_ratio=0.05_varing_n}).
By contrast, \emph{boot-$i$}, \emph{boot-$ij_r$}, and \emph{boot-$ij_s$} overestimated $\sigma_r^2$ (by about $+20\%$ to $+25\%$) and underestimated $\sigma_L^2$ under the same configuration.
 This advantage of \emph{boot-$j_r$} and \emph{boot-$j_s$} is also observed when $k$ or $n$ varies with the other fixed at $5$ (Table~\ref{tab:PE_n=5_ratio=0.05_varing_k}) and across variance ratios with $(k,n)=(5,5)$ (Table~\ref{tab:PE_n=5_k=5_varing_ratio}).

Next, we examined the bootstrap means in (\ref{eq:8}) and the proposed adjusted point estimators in (\ref{eq:11})–(\ref{eq:13}).
Three sets of comparisons were conducted: (i) $k = 5$ with varying $n$ (Table~\ref{tab:PE_k=5_ratio=0.05_varing_n}), (ii) $n = 5$ with varying $k$ (Table~\ref{tab:PE_n=5_ratio=0.05_varing_k}), and (iii) $(k, n) = (5, 5)$ with varying $\sigma^2_L / \sigma^2_r$ (Table~\ref{tab:PE_n=5_k=5_varing_ratio}).
As a working criterion, we regarded estimates as sufficiently close to the truth if they fell within $\pm 0.01$ of the true values.
For $k=5$, \emph{boot-$i$} produced bootstrap means close to the true $\sigma^2_r$, whereas the other methods tended to underestimate this variance; \emph{boot-$i$} again showed underestimation for $\sigma^2_L$.
The other methods overestimated $\sigma^2_L$ for small $n$, but the bias reversed to underestimation as $n$ increased.
Among the adjusted estimators, \emph{boot-$j_r$} and \emph{boot-$j_s$} most frequently delivered estimates within $\pm0.01$ of the true values, with only a few exceptions.

Even in the relatively small configuration $(k,n)=(5,5)$, both \emph{boot-$j_r$} and \emph{boot-$j_s$} produced estimates very close to the true values for both $\sigma^2_r$ ($\hat{\sigma}_{r:\mathrm{ad}}^2 \approx 1.01$) and $\sigma^2_L$ ($\hat{\sigma}_{L:\mathrm{ad}}^2 \approx 0.50$).
By contrast, \emph{boot-$i$}, \emph{boot-$ij_r$}, and \emph{boot-$ij_s$} overestimated $\sigma^2_r$ by roughly $25\%$ and underestimated $\sigma^2_L$ under the same setting.
Performance improved as $n$ increased, eventually approaching that of the \emph{boot-$j_r$} estimators (e.g., $n=50$; Table~\ref{tab:PE_k=5_ratio=0.05_varing_n}).
In extremely small samples (e.g., $(k,n)=(3,3)$), the adjustment factors (such as $k/(k-1)=1.5$) were too strong and increased estimation errors (Table~\ref{tab:PE_n_k_ratio_extreme_val}).
Furthermore, when the ratio $\sigma_L^2/\sigma_r^2$ was small (Table~\ref{tab:PE_n=5_k=5_varing_ratio}), \emph{boot-$i$}, \emph{boot-$ij_r$}, and \emph{boot-$ij_s$} exhibited even more noticeable underestimation of $\sigma^2_L$.
Overall, \emph{boot-$j_r$} and \emph{boot-$j_s$} yielded the most stable point estimates across variations in sample size and in the ratio of the two variance components.

 \subsubsection{Standard error (SE)}
 \label{sec:stand-error}

 A consistent pattern was observed across methods (Tables~\ref{tab:PE_k=5_ratio=0.05_varing_n}--\ref{tab:PE_n=5_k=5_varing_ratio}).
Relative to the reference ANOVA SEs (Eqs.~(\ref{eq:5})--(\ref{eq:7})), the within-laboratory resampling strategies (\emph{boot-$j_r$} and \emph{boot-$j_s$}) generally yielded smaller or comparable SEs, whereas the two-stage and double resampling strategies (\emph{boot-$ij_r$} and \emph{boot-$ij_s$}) tended to yield larger SEs.
The \emph{boot-$i$} results were typically close to the ANOVA SEs.
This qualitative pattern held across changes in sample size and in the variance ratio.
Quantitatively, for $\sigma_r^2$ with $(k,n)=(5,5)$ and $\sigma_L^2/\sigma_r^2=0.5$, the ANOVA SE is $0.302$, while \emph{boot-$j_r$} and \emph{boot-$j_s$} give $0.285$ and $0.313$, respectively; by contrast, \emph{boot-$ij_r$} and \emph{boot-$ij_s$} yield substantially larger SEs ($0.476$ and $0.591$) (Table~\ref{tab:PE_k=5_ratio=0.05_varing_n}).
Similar patterns appear when varying $k$ or $n$ (Table~\ref{tab:PE_n=5_ratio=0.05_varing_k}) and across variance ratios (Table~\ref{tab:PE_n=5_k=5_varing_ratio}).
This systematic difference in SE is reflected in the CI behavior observed in Section~\ref{sec:conf-interv}: smaller SEs lead to narrower intervals and undercoverage, whereas larger SEs induce more conservative coverage.

\subsubsection{Confidence intervals (CIs)}
\label{sec:conf-interv}

We primarily evaluated CI performance using the CP of $95\% CIs$.
When the bootstrap mean was used to construct CIs (Table~\ref{tab:CI_bootstrap_mean}), cases in which CP achieved the nominal 95\% level were rare, regardless of the resampling scheme.
In contrast, CPs generally improved when adjusted estimators were used (Table~\ref{tab:CI_adjusted}), which confirms the advantage of the adjusted approach.

Using the adjusted estimators (Table~\ref{tab:CI_adjusted}), the choice of resampling method directly affected CPs.
Methods such as \emph{boot-$j_r$} and \emph{boot-$j_s$}, which tend to yield smaller SEs, produced intervals that were too narrow and led to undercoverage (e.g., for $\sigma^2_R$ at $(k,n)=(5,5)$ and $\sigma^2_L/\sigma^2_r=0.5$, CP was $0.774$--$0.882$ depending on the CI method).
Conversely, \emph{boot-$ij_r$} and \emph{boot-$ij_s$}, which produce larger SEs, yielded wider intervals and CPs at or above $95\%$ (e.g., $0.961$--$0.968$ for $\sigma^2_R$ under the same configuration when paired with BCa), and indiced conservative yet reliable intervals.

Finally, Table~\ref{tab:CI_n_k_ratio_extreme_val} examines the impact of extreme combinations of $(k,n)$ and $\sigma^2_L/\sigma^2_r$.
At the minimum sample size $(k,n)=(3,3)$---a setting in which even approximate methods become unstable---\emph{boot-$ij_r$} maintained the most stable CP (around $90\%$).
When $\sigma^2_L/\sigma^2_r$ increased, CPs for $\sigma^2_L$ and $\sigma^2_R$ decreased substantially; this tendency persisted even at $(k,n)=(50,50)$, where the CP of \emph{boot-$ij_r$} reached only about $92\%$--$94\%$.
The relative robustness of \emph{boot-$ij_r$} with BCa under extreme designs is also evident in Table~\ref{tab:CI_n_k_ratio_extreme_val}:
at $(k,n)=(3,3)$, it maintained the most stable coverage for $\sigma^2_R \approx 0.96)$, whereas \emph{boot-$j_r$}-based methods showed marked undercoverage for $\sigma^2_L$ and $\sigma^2_R$ when $\sigma^2_L/\sigma^2_r$ increased (see the $(k,n) = (50,50)$ cases with ratio $2.00$).

\subsubsection{Overall conclustion for the results}
\label{sec:over-concl-for-resul}

For point estimation, the adjusted \emph{boot-$j_r$} and \emph{boot-$j_s$} are preferred in small to moderate samples due to their low bias relative to the true values (Tables~\ref{tab:PE_k=5_ratio=0.05_varing_n}--\ref{tab:PE_n_k_ratio_extreme_val}).
For interval estimation, the adjusted \emph{boot-$ij_r$} with BCa is the safest and attained near-nominal or conservative coverage across a broad range of designs (Tables~\ref{tab:CI_bootstrap_mean}--\ref{tab:CI_n_k_ratio_extreme_val}).

\section{Case study}
\label{sec:case-stad}

\subsection{Real example}
\label{sec:real-example}

We applied the five resampling schemes (Section~\ref{sec:resampl-schem}) to a practical dataset from ISO 5725-4~\cite{iso2020ISO572542025}, namely the manganese content in iron ore interlaboratory study.
For each scheme, we estimated the three precision variances, $\sigma^2_r$, $\sigma^2_L$, and $\sigma^2_R=\sigma^2_r+\sigma^2_L$, using the same estimators and notations as in the simulation-study section.
The number of bootstrap replicates was fixed at $M=1000$, as matching our simulation study, and for each method we reported (i) the bootstrap mean (Eq.~(\ref{eq:8})), (ii) the bias-corrected estimator (Eq.~(\ref{eq:9})), and (iii) the adjusted estimator based on Eqs.~(\ref{eq:11})--~(\ref{eq:13}).
As a non-resampling benchmark, we also computed the ANOVA estimators (Eqs. (\ref{eq:4})-- (\ref{eq:7}).

\subsection{Result of the example}
\label{sec:result-exmaple}

Because the variance estimates in this dataset are numerically small, all reported values in Tables~\ref{tab:ex_point_estimates} and \ref{tab:adjusted_CIs_practical_swapped} are scaled by $10^{7}$ for readability.

Table~\ref{tab:ex_point_estimates} reports point estimates for the practical dataset.
For comparison, ANOVA estimates without resampling are included.
Since the true values are unknown, accuracy cannot be assessed; however, these results are informative about the behavior of the estimators in practice.
For bootstrap means,  \emph{boot-$i$} for $\sigma^2_r$, and \emph{boot-$j_r$} and \emph{boot-$j_s$} for $\sigma^2_R$ are broadly consistent with ANOVA.
For bias-corrected estimators, the same schemes remain consistent with ANOVA.
Using the adjusted formulas, \emph{boot-$j_r$} and \emph{boot-$j_s$} for all the variance components and \emph{boot-$ij_r$} for $\sigma^2_R$ are broadly in agreement with ANOVA, which is consistent with the results of the simulation study.
Moreover, for \emph{boot-$i$} for $\sigma^2_L$ and \emph{boot-$j_r$}/\emph{boot-$j$} for $\sigma^2_R$, the bootstrap mean and the bias-corrected estimates are nearly identical, which suggests that the bias is negligible in this setting.
Estimated SEs follow the same ordering as in the results of the simulation study: smaller for \emph{boot-$j_r$}/\emph{boot-$j$} , intermediate for  \emph{boot-$i$}/ANOVA, and larger for \emph{boot-$ij_r$}/\emph{boot-$ij_s$}.

Table~\ref{tab:adjusted_CIs_practical_swapped} shows the $95\%$ CIs.
A notable feature is that the Moriguchi-type approximate upper limit for $\sigma^2_L$ reaches $128.30$, whereas resampling-based upper limits are smaller; a similar pattern holds for $\sigma^2_R$.
Overall, results most similar to the approximate intervals are limited to  \emph{boot-$j_r$} and \emph{boot-$j_s$} for $\sigma^2_r$; for other components, resampling intervals tend to be relatively tighter on the upper side.

\section{Discussion}
\label{sec:disc}

This simulation study indicates that the preferred approach differs between point estimation and CI construction.
In brief, for point estimation, the adjusted \emph{boot-$j_r$} and \emph{boot-$j_s$} estimators yielded the most accurate and stable results across a broad range of scenarios, whereas for CI construction, the adjusted \emph{boot-$ij_r$} estimator, particularly when coupled with BCa, most reliably achieved or exceeded the nominal coverage level.

For point estimation, the dominant criteria are low bias and stability.
Across small-sample settings such as $(k,n)=(5,5)$ and across heterogeneous conditions induced by varying the variance ratio $\sigma_L^2/\sigma_r^2$, the adjusted \emph{boot-$j_r$} and \emph{boot-$j_s$} estimators consistently produced estimates closest to the true values.
By contrast, \emph{boot-$i$}, \emph{boot-$ij_r$}, and \emph{boot-$ij_s$} exhibited instability in extremely small samples, e.g., $(k,n)=(3,3)$, where the adjustment tended to over-correct and degrade accuracy.
These findings suggest that, in the small-sample cases that frequently arise in practice, \emph{boot-$j_r$} and \emph{boot-$j_s$} schemes are preferable to \emph{boot-$i$}, \emph{boot-$ij_r$}, and \emph{boot-$ij_s$} for point estimation.

For CI construction, the critical requirement is that the CP meets the nominal level ($95\%$ in this study), and CI construction therefore requires different considerations than point estimation.
Methods that excelled in point estimation (\emph{boot-$j_r$} and \emph{boot-$j_s$}) tended to underestimate the SE, which produced narrower intervals and sub-nominal coverage (undercoverage).
Conversely, \emph{boot-$ij_r$} scheme tended to slightly overestimate SE; although this inflates interval width, the resulting CPs were at or above $95\%$ (conservative).
For example, for $\sigma_R^2$ with $(k,n)=(5,5)$ and $\sigma_L^2/\sigma_r^2=0.5$, CIs based on adjusted \emph{boot-$j_r$}/\emph{boot-$j_s$} achieved CPs of $0.774$–$0.882$, whereas adjusted \emph{boot-$ij_r$} with BCa attained $0.961$–$0.968$ (Table~\ref{tab:CI_adjusted}). 
In practice, undercoverage (not including the true value) is typically treated as a more serious error than comparable over-conservatism.
Accordingly, for CI construction, the adjusted \emph{boot-$ij_r$} estimator with BCa is, in general, the safer and more reliable choice.

Regarding the real example, in point estimation, we observed consistency with ANOVA for \emph{boot-$i$} on $\sigma^2_r$ and  \emph{boot-$j_r$}/\emph{boot-$j_s$} on $\sigma^2_R$;
when the adjusted formulas were applied, this agreement extended to \emph{boot-$j_r$}/\emph{boot-$j_s$} for all variance components and to \emph{boot-$ij_r$} for $\sigma^2_R$.
  This mirrors the simulation finding that \emph{boot-$j_r$}/\emph{boot-$j_s$} provide stable point estimates with relatively small SEs.
  The near-identity between bootstrap means and bias-corrected estimates for certain components suggests negligible bias.
  The SE ordering (\emph{boot-$j_r$}/\emph{boot-$j_s$} $<$ \emph{boot-$i$}/ANOVA $<$ \emph{boot-$ij_r$}/\emph{boot-$ij_s$}) also matches the simulations, which implies that resampling-based CIs may be tighter (less conservative) than approximate ones for some components.
  Indeed, the Moriguchi-type upper limits are substantially larger than those from resampling, particularly for $\sigma^2_L$ and similarly for $\sigma^2_R$ in this dataset.
  Because the true values are unknown, we refrain from judging accuracy;
  nevertheless, the example clarifies how methods behave in practice and helps guide reporting:
  we recommend using \emph{boot-$j_r$} or \emph{boot-$j_s$} for stable point estimation, and then selecting the CI method depending on whether a conservative or a tighter interval is desired.

 Finally, the study explains under what conditions performance declines and the reasons for it, and provides practical guidance for method selection.
 When the variance ratio $\sigma_L^2/\sigma_r^2$ exceeded $1.0$, CPs tended to fall below the nominal level even for CIs based on adjusted \emph{boot-$ij_r$} with BCa, and
suggests that when the between-laboratory component dominates, bootstrap-based approaches become less reliable.
Moreover, when $k$ and $n$ were highly imbalanced, one large and the other small, CP sometimes became overly conservative.
These patterns highlight two practical implications.
First, CI performance depends not only on sample size but also on $\sigma_L^2/\sigma_r^2$.
Second, imbalance between $k$ and $n$ can degrade CP, even when the overall information is large.

In summary, we recommend adjusted \emph{boot-$j_r$} and \emph{boot-$j_s$} for point estimation and adjusted \emph{boot-$ij_r$} estimator with BCa for CI construction.
Future work should address (i) procedures that maintain nominal coverage when $\sigma_L^2/\sigma_r^2>1.0$, and (ii) designs or adjustments that mitigate the effect of severe $k$–$n$ imbalance on CI performance.

\section*{Acknowledgements}
During the preparation of this manuscript, the authors used \textit{Grammarly (version 1.134.1.0)} and \textit{Microsoft 365 Copilot} to assist with English grammar, style, and \LaTeX{} formatting.
The authors reviewed and edited the manuscript and take full responsibility for the content.

\clearpage

\bibliographystyle{siam}
\bibliography{ResamplePrecEst}

\clearpage

\begin{landscape}
\begin{table}[htbp]\centering
  \small
\caption{Bootstrap means, adjusted point estimators, and their standard errors for $k=5$ and $\sigma_L^2/\sigma_r^2 = 0.05$ with varying $n$.}
\label{tab:PE_k=5_ratio=0.05_varing_n}
\begin{tabular}{ccc|rrrrrr|rrrrrr}
\hline
  $k$ & $n$ &  & 
  $\sigma_{r:(\cdot)}^2$ & $\mathrm{SE}\!\left(\sigma_{r:(\cdot)}^2\right)$ &
  $\sigma_{L:(\cdot)}^2$ & $\mathrm{SE}\!\left(\sigma_{L:(\cdot)}^2\right)$ &
  $\sigma_{R:(\cdot)}^2$ & $\mathrm{SE}\!\left(\sigma_{R:(\cdot)}^2\right)$ &
  $\sigma_{r:\mathrm{ad}}^2$ & $\mathrm{SE}\!\left(\sigma_{r:\mathrm{ad}}^2\right)$ &
  $\sigma_{L:\mathrm{ad}}^2$ & $\mathrm{SE}\!\left(\sigma_{L:\mathrm{ad}}^2\right)$ &
  $\sigma_{R:\mathrm{ad}}^2$ & $\mathrm{SE}\!\left(\sigma_{R:\mathrm{ad}}^2\right)$ \\
\hline
 &  & True value 
 & $1.00$ &  & $0.50$ &  & $1.50$ &  &  &  &  &  &  &  \\
\hline
5 & 3 & ANOVA 
& $1.006$ & $0.411$ & $0.497$ & $0.514$ & $1.503$ & $0.569$
&  &  &  &  &  &  \\
 &   & \emph{boot-$i$}    
& $1.004$ & $0.347$ & $0.317$ & $0.382$ & $1.321$ & $0.445$
& $1.254$ & $0.434$ & $0.396$ & $0.478$ & $1.651$ & $0.556$ \\
 &   & \emph{boot-$j_r$}  
& $0.665$ & $0.268$ & $0.825$ & $0.456$ & $1.490$ & $0.417$
& $0.998$ & $0.402$ & $0.492$ & $0.529$ & $1.490$ & $0.417$ \\
 &   & \emph{boot-$j_s$} 
& $0.676$ & $0.354$ & $0.842$ & $0.505$ & $1.518$ & $0.401$
& $1.014$ & $0.531$ & $0.504$ & $0.630$ & $1.518$ & $0.401$ \\
 &   & \emph{boot-$ij_r$}  
& $0.662$ & $0.355$ & $0.677$ & $0.583$ & $1.339$ & $0.610$
& $1.241$ & $0.665$ & $0.432$ & $0.825$ & $1.673$ & $0.763$ \\
 &   &\emph{boot-$ij_s$}
& $0.664$ & $0.474$ & $0.610$ & $0.672$ & $1.274$ & $0.660$
& $1.245$ & $0.889$ & $0.348$ & $1.005$ & $1.593$ & $0.825$ \\
\hline
5 & 5 & ANOVA
& $1.000$ & $0.302$ & $0.499$ & $0.412$ & $1.499$ & $0.491$
&  &  &  &  &  &  \\
 &   & \emph{boot-$i$}    
& $0.997$ & $0.249$ & $0.363$ & $0.311$ & $1.360$ & $0.381$
& $1.246$ & $0.312$ & $0.454$ & $0.388$ & $1.700$ & $0.477$ \\
 &   & \emph{boot-$j_r$}  
& $0.806$ & $0.228$ & $0.707$ & $0.345$ & $1.513$ & $0.370$
& $1.008$ & $0.285$ & $0.505$ & $0.363$ & $1.513$ & $0.370$ \\
 &   & \emph{boot-$j_s$} 
& $0.809$ & $0.251$ & $0.704$ & $0.360$ & $1.513$ & $0.356$
& $1.011$ & $0.313$ & $0.502$ & $0.388$ & $1.513$ & $0.356$ \\
 &   & \emph{boot-$ij_r$}  
& $0.791$ & $0.305$ & $0.555$ & $0.444$ & $1.346$ & $0.512$
& $1.236$ & $0.476$ & $0.447$ & $0.581$ & $1.683$ & $0.640$ \\
 &   &\emph{boot-$ij_s$}
& $0.794$ & $0.378$ & $0.526$ & $0.504$ & $1.320$ & $0.573$
& $1.241$ & $0.591$ & $0.409$ & $0.671$ & $1.651$ & $0.716$ \\
\hline
5 & 10 & ANOVA
& $1.003$ & $0.207$ & $0.493$ & $0.344$ & $1.496$ & $0.407$
&  &  &  &  &  &  \\
 &    & \emph{boot-$i$}    
& $1.008$ & $0.175$ & $0.379$ & $0.261$ & $1.388$ & $0.321$
& $1.260$ & $0.219$ & $0.474$ & $0.326$ & $1.735$ & $0.401$ \\
 &    & \emph{boot-$j_r$}  
& $0.899$ & $0.177$ & $0.603$ & $0.230$ & $1.503$ & $0.278$
& $0.999$ & $0.196$ & $0.503$ & $0.233$ & $1.503$ & $0.278$ \\
 &    & \emph{boot-$j_s$} 
& $0.894$ & $0.179$ & $0.588$ & $0.228$ & $1.482$ & $0.271$
& $0.994$ & $0.199$ & $0.489$ & $0.232$ & $1.482$ & $0.271$ \\
 &    & \emph{boot-$ij_r$}  
& $0.908$ & $0.242$ & $0.484$ & $0.342$ & $1.392$ & $0.419$
& $1.262$ & $0.336$ & $0.479$ & $0.432$ & $1.741$ & $0.524$ \\
 &    &\emph{boot-$ij_s$}
& $0.905$ & $0.293$ & $0.445$ & $0.363$ & $1.350$ & $0.456$
& $1.257$ & $0.407$ & $0.431$ & $0.462$ & $1.688$ & $0.570$ \\
\hline
5 & 50 & ANOVA
& $1.001$ & $0.090$ & $0.499$ & $0.299$ & $1.499$ & $0.319$
&  &  &  &  &  &  \\
 &    & boot-i' 
& $1.400$ & $0.110$ & $-0.004$ & $0.017$ & $1.396$ & $0.109$
& $1.750$ & $0.138$ & $-0.005$ & $0.021$ & $1.745$ & $0.136$ \\
 &    & \emph{boot-$j_r$}  
& $0.978$ & $0.087$ & $0.517$ & $0.095$ & $1.496$ & $0.130$
& $0.998$ & $0.089$ & $0.497$ & $0.096$ & $1.496$ & $0.130$ \\
 &    & \emph{boot-$j_s$} 
& $0.978$ & $0.087$ & $0.505$ & $0.094$ & $1.482$ & $0.129$
& $0.998$ & $0.088$ & $0.485$ & $0.095$ & $1.482$ & $0.129$ \\
 &    & \emph{boot-$ij_r$}  
& $0.981$ & $0.117$ & $0.411$ & $0.241$ & $1.392$ & $0.276$
& $1.251$ & $0.149$ & $0.489$ & $0.302$ & $1.740$ & $0.345$ \\
 &    &\emph{boot-$ij_s$}
& $0.982$ & $0.141$ & $0.418$ & $0.256$ & $1.400$ & $0.300$
& $1.253$ & $0.179$ & $0.497$ & $0.320$ & $1.750$ & $0.375$ \\
\hline
\end{tabular}
\end{table}
\end{landscape}

\begin{landscape}
\begin{table}[htbp]\centering
\small
\caption{Bootstrap means, adjusted point estimators, and their standard errors for $n=5$ and $\sigma_L^2/\sigma_r^2 = 0.05$ with varying $k$.}
\label{tab:PE_n=5_ratio=0.05_varing_k}
\begin{tabular}{ccc|rrrrrr|rrrrrr}
\hline
  $k$ & $n$ &  & 
  $\sigma_{r:(\cdot)}^2$ & $\mathrm{SE}\!\left(\sigma_{r:(\cdot)}^2\right)$ &
  $\sigma_{L:(\cdot)}^2$ & $\mathrm{SE}\!\left(\sigma_{L:(\cdot)}^2\right)$ &
  $\sigma_{R:(\cdot)}^2$ & $\mathrm{SE}\!\left(\sigma_{R:(\cdot)}^2\right)$ &
  $\sigma_{r:\mathrm{ad}}^2$ & $\mathrm{SE}\!\left(\sigma_{r:\mathrm{ad}}^2\right)$ &
  $\sigma_{L:\mathrm{ad}}^2$ & $\mathrm{SE}\!\left(\sigma_{L:\mathrm{ad}}^2\right)$ &
  $\sigma_{R:\mathrm{ad}}^2$ & $\mathrm{SE}\!\left(\sigma_{R:\mathrm{ad}}^2\right)$ \\
\hline
 &  & True value 
 & $1.00$ &  & $0.50$ &  & $1.50$ &  &  &  &  &  &  &  \\
\hline
3 & 5 & ANOVA
& $0.998$ & $0.377$ & $0.502$ & $0.512$ & $1.500$ & $0.624$
&  &  &  &  &  &  \\
 &   & \emph{boot-$i$}
& $1.013$ & $0.284$ & $0.250$ & $0.332$ & $1.263$ & $0.433$
& $1.520$ & $0.425$ & $0.375$ & $0.498$ & $1.895$ & $0.649$ \\
 &   & \emph{boot-$j_r$}
& $0.810$ & $0.286$ & $0.705$ & $0.468$ & $1.515$ & $0.495$
& $1.013$ & $0.358$ & $0.502$ & $0.491$ & $1.515$ & $0.495$ \\
 &   & \emph{boot-$j_s$}
& $0.807$ & $0.297$ & $0.699$ & $0.452$ & $1.505$ & $0.467$
& $1.009$ & $0.371$ & $0.497$ & $0.481$ & $1.505$ & $0.467$ \\
 &   & \emph{boot-$ij_r$}
& $0.781$ & $0.361$ & $0.474$ & $0.537$ & $1.255$ & $0.622$
& $1.465$ & $0.677$ & $0.418$ & $0.845$ & $1.883$ & $0.934$ \\
 &   &\emph{boot-$ij_s$}
& $0.783$ & $0.438$ & $0.429$ & $0.580$ & $1.212$ & $0.679$
& $1.468$ & $0.821$ & $0.350$ & $0.929$ & $1.818$ & $1.019$ \\
\hline
$5$ & $5$ & ANOVA
& $1.000$ & $0.302$ & $0.499$ & $0.412$ & $1.499$ & $0.491$
&  &  &  &  &  &  \\
 &   & \emph{boot-$i$}
& $0.997$ & $0.249$ & $0.363$ & $0.311$ & $1.360$ & $0.381$
& $1.246$ & $0.312$ & $0.454$ & $0.388$ & $1.700$ & $0.477$ \\
 &   & \emph{boot-$j_r$}
& $0.806$ & $0.228$ & $0.707$ & $0.345$ & $1.513$ & $0.370$
& $1.008$ & $0.285$ & $0.505$ & $0.363$ & $1.513$ & $0.370$ \\
 &   & \emph{boot-$j_s$}
& $0.809$ & $0.251$ & $0.704$ & $0.360$ & $1.513$ & $0.356$
& $1.011$ & $0.313$ & $0.502$ & $0.388$ & $1.513$ & $0.356$ \\
 &   & \emph{boot-$ij_r$}
& $0.791$ & $0.305$ & $0.555$ & $0.444$ & $1.346$ & $0.512$
& $1.236$ & $0.476$ & $0.447$ & $0.581$ & $1.683$ & $0.640$ \\
 &   &\emph{boot-$ij_s$}
& $0.794$ & $0.378$ & $0.526$ & $0.504$ & $1.320$ & $0.573$
& $1.241$ & $0.591$ & $0.409$ & $0.671$ & $1.651$ & $0.716$ \\
\hline
$10$ & $5$ & ANOVA
& $0.997$ & $0.218$ & $0.494$ & $0.300$ & $1.491$ & $0.350$
&  &  &  &  &  &  \\
 &    & \emph{boot-$i$}
& $1.002$ & $0.198$ & $0.428$ & $0.257$ & $1.429$ & $0.305$
& $1.113$ & $0.220$ & $0.475$ & $0.286$ & $1.588$ & $0.338$ \\
 &    & \emph{boot-$j_r$}
& $0.791$ & $0.163$ & $0.698$ & $0.235$ & $1.489$ & $0.253$
& $0.989$ & $0.203$ & $0.500$ & $0.248$ & $1.489$ & $0.253$ \\
 &    & \emph{boot-$j_s$}
& $0.792$ & $0.198$ & $0.690$ & $0.256$ & $1.482$ & $0.236$
& $0.990$ & $0.248$ & $0.492$ & $0.284$ & $1.482$ & $0.236$ \\
 &    & \emph{boot-$ij_r$}
& $0.800$ & $0.231$ & $0.622$ & $0.342$ & $1.421$ & $0.394$
& $1.111$ & $0.321$ & $0.469$ & $0.395$ & $1.579$ & $0.438$ \\
 &    &\emph{boot-$ij_s$}
& $0.802$ & $0.301$ & $0.602$ & $0.405$ & $1.404$ & $0.446$
& $1.114$ & $0.418$ & $0.446$ & $0.479$ & $1.560$ & $0.496$ \\
\hline
$50$ & $5$ & ANOVA
& $1.000$ & $0.100$ & $0.498$ & $0.140$ & $1.499$ & $0.160$
&  &  &  &  &  &  \\
 &    & \emph{boot-$i$}
& $1.001$ & $0.097$ & $0.484$ & $0.135$ & $1.485$ & $0.155$
& $1.022$ & $0.099$ & $0.494$ & $0.138$ & $1.516$ & $0.159$ \\
 &    & \emph{boot-$j_r$}
& $0.799$ & $0.075$ & $0.696$ & $0.104$ & $1.494$ & $0.113$
& $0.999$ & $0.094$ & $0.496$ & $0.110$ & $1.494$ & $0.113$ \\
 &    & \emph{boot-$j_s$}
& $0.798$ & $0.145$ & $0.692$ & $0.162$ & $1.490$ & $0.107$
& $0.998$ & $0.181$ & $0.492$ & $0.190$ & $1.490$ & $0.107$ \\
 &    & \emph{boot-$ij_r$}
& $0.800$ & $0.108$ & $0.692$ & $0.170$ & $1.492$ & $0.193$
& $1.021$ & $0.138$ & $0.502$ & $0.179$ & $1.523$ & $0.196$ \\
 &    &\emph{boot-$ij_s$}
& $0.799$ & $0.180$ & $0.681$ & $0.231$ & $1.481$ & $0.220$
& $1.019$ & $0.230$ & $0.491$ & $0.260$ & $1.511$ & $0.225$ \\
\hline
\end{tabular}
\end{table}
\end{landscape}

\begin{landscape}
\begin{table}[htbp]\centering
\small
\caption{Bootstrap means, adjusted point estimators, and their standard errors for $(n,k)=(5,5)$ with varying $\sigma_L^2/\sigma_r^2$.}
\label{tab:PE_n=5_k=5_varing_ratio}
\begin{tabular}{ccc|rrrrrr|rrrrrr}
\hline
  $k$ & $\sigma_L^2 / \sigma_r^2$ &  &
  $\sigma_{r:(\cdot)}^2$ & $\mathrm{SE}\!\left(\sigma_{r:(\cdot)}^2\right)$ &
  $\sigma_{L:(\cdot)}^2$ & $\mathrm{SE}\!\left(\sigma_{L:(\cdot)}^2\right)$ &
  $\sigma_{R:(\cdot)}^2$ & $\mathrm{SE}\!\left(\sigma_{R:(\cdot)}^2\right)$ &
  $\sigma_{r:\mathrm{ad}}^2$ & $\mathrm{SE}\!\left(\sigma_{r:\mathrm{ad}}^2\right)$ &
  $\sigma_{L:\mathrm{ad}}^2$ & $\mathrm{SE}\!\left(\sigma_{L:\mathrm{ad}}^2\right)$ &
  $\sigma_{R:\mathrm{ad}}^2$ & $\mathrm{SE}\!\left(\sigma_{R:\mathrm{ad}}^2\right)$ \\
\hline
 &  & True value 
 & $1.00$ &  & $0.25$ &  & $1.25$ &  &  &  &  &  &  &  \\
\hline
$5$ & $0.25$ & ANOVA
& $1.006$ & $0.303$ & $0.249$ & $0.272$ & $1.255$ & $0.373$
&  &  &  &  &  &  \\
 &        & \emph{boot-$i$}
& $0.993$ & $0.257$ & $0.162$ & $0.203$ & $1.155$ & $0.302$
& $1.242$ & $0.321$ & $0.202$ & $0.254$ & $1.444$ & $0.378$ \\
 &        & \emph{boot-$j_r$}
& $0.803$ & $0.229$ & $0.455$ & $0.289$ & $1.259$ & $0.317$
& $1.004$ & $0.287$ & $0.254$ & $0.311$ & $1.259$ & $0.317$ \\
 &        &\emph{boot-$j_s$}
& $0.805$ & $0.249$ & $0.445$ & $0.305$ & $1.250$ & $0.305$
& $1.006$ & $0.312$ & $0.243$ & $0.336$ & $1.250$ & $0.305$ \\
 &        & \emph{boot-$ij_r$}
& $0.807$ & $0.311$ & $0.358$ & $0.337$ & $1.165$ & $0.420$
& $1.261$ & $0.485$ & $0.195$ & $0.454$ & $1.456$ & $0.525$ \\
 &        &\emph{boot-$ij_s$}
& $0.808$ & $0.386$ & $0.323$ & $0.391$ & $1.131$ & $0.477$
& $1.263$ & $0.603$ & $0.151$ & $0.540$ & $1.414$ & $0.596$ \\
\hline
$5$ & $0.50$ & ANOVA
& $1.000$ & $0.302$ & $0.499$ & $0.412$ & $1.499$ & $0.491$
&  &  &  &  &  &  \\
 &        & \emph{boot-$i$}
& $0.997$ & $0.249$ & $0.363$ & $0.311$ & $1.360$ & $0.381$
& $1.246$ & $0.312$ & $0.454$ & $0.388$ & $1.700$ & $0.477$ \\
 &        & \emph{boot-$j_r$}
& $0.806$ & $0.228$ & $0.707$ & $0.345$ & $1.513$ & $0.370$
& $1.008$ & $0.285$ & $0.505$ & $0.363$ & $1.513$ & $0.370$ \\
 &        &\emph{boot-$j_s$}
& $0.809$ & $0.251$ & $0.704$ & $0.360$ & $1.513$ & $0.356$
& $1.011$ & $0.313$ & $0.502$ & $0.388$ & $1.513$ & $0.356$ \\
 &        & \emph{boot-$ij_r$}
& $0.791$ & $0.305$ & $0.555$ & $0.444$ & $1.346$ & $0.512$
& $1.236$ & $0.476$ & $0.447$ & $0.581$ & $1.683$ & $0.640$ \\
 &        &\emph{boot-$ij_s$}
& $0.794$ & $0.378$ & $0.526$ & $0.504$ & $1.320$ & $0.573$
& $1.241$ & $0.591$ & $0.409$ & $0.671$ & $1.651$ & $0.716$ \\
\hline
$5$ & $1.00$ & ANOVA
& $0.995$ & $0.300$ & $0.981$ & $0.686$ & $1.976$ & $0.743$
&  &  &  &  &  &  \\
 &        & \emph{boot-$i$}
& $0.987$ & $0.250$ & $0.751$ & $0.520$ & $1.737$ & $0.576$
& $1.233$ & $0.312$ & $0.938$ & $0.650$ & $2.172$ & $0.720$ \\
 &        & \emph{boot-$j_r$}
& $0.798$ & $0.225$ & $1.158$ & $0.420$ & $1.957$ & $0.438$
& $0.998$ & $0.281$ & $0.959$ & $0.436$ & $1.957$ & $0.438$ \\
 &        &\emph{boot-$j_s$}
& $0.809$ & $0.249$ & $1.143$ & $0.430$ & $1.952$ & $0.436$
& $1.012$ & $0.311$ & $0.941$ & $0.452$ & $1.952$ & $0.436$ \\
 &        & \emph{boot-$ij_r$}
& $0.797$ & $0.305$ & $0.938$ & $0.656$ & $1.735$ & $0.709$
& $1.245$ & $0.477$ & $0.924$ & $0.839$ & $2.169$ & $0.886$ \\
 &        &\emph{boot-$ij_s$}
& $0.796$ & $0.384$ & $0.912$ & $0.717$ & $1.708$ & $0.771$
& $1.244$ & $0.600$ & $0.891$ & $0.929$ & $2.135$ & $0.963$ \\
\hline
$5$ & $2.00$ & ANOVA
& $0.997$ & $0.301$ & $1.992$ & $1.268$ & $2.990$ & $1.303$
&  &  &  &  &  &  \\
 &        & \emph{boot-$i$}
& $0.989$ & $0.249$ & $1.525$ & $0.927$ & $2.515$ & $0.961$
& $1.237$ & $0.312$ & $1.906$ & $1.159$ & $3.143$ & $1.202$ \\
 &        & \emph{boot-$j_r$}
& $0.801$ & $0.227$ & $2.194$ & $0.565$ & $2.994$ & $0.580$
& $1.001$ & $0.283$ & $1.993$ & $0.577$ & $2.994$ & $0.580$ \\
 &        &\emph{boot-$j_s$}
& $0.810$ & $0.251$ & $2.206$ & $0.561$ & $3.015$ & $0.560$
& $1.012$ & $0.314$ & $2.003$ & $0.581$ & $3.015$ & $0.560$ \\
 &        & \emph{boot-$ij_r$}
& $0.789$ & $0.299$ & $1.768$ & $1.103$ & $2.556$ & $1.137$
& $1.232$ & $0.468$ & $1.963$ & $1.392$ & $3.196$ & $1.421$ \\
 &        &\emph{boot-$ij_s$}
& $0.802$ & $0.380$ & $1.741$ & $1.174$ & $2.543$ & $1.207$
& $1.254$ & $0.594$ & $1.926$ & $1.491$ & $3.179$ & $1.508$ \\
\hline
\end{tabular}
\end{table}
\end{landscape}

\begin{landscape}
\begin{table}[htbp]\centering
\small
\caption{Bootstrap means, adjusted point estimators, and their standard errors under extreme values of $n$, $k$, and $\sigma_L^2/\sigma_r^2$.}
\label{tab:PE_n_k_ratio_extreme_val}
\begin{tabular}{cccc|rrrrrr|rrrrrr}
\hline
  $k$ & $n$ & $\sigma_L^2/ \sigma_r^2$ & &
  $\sigma_{r:(\cdot)}^2$ & $\mathrm{SE}\!\left( \sigma_{r:(\cdot)}^2 \right)$ &
  $\sigma_{L:(\cdot)}^2$ & $\mathrm{SE}\!\left( \sigma_{L:(\cdot)}^2 \right)$ &
  $\sigma_{R:(\cdot)}^2$ & $\mathrm{SE}\!\left( \sigma_{R:(\cdot)}^2 \right)$ &
  $\sigma_{r:\mathrm{ad}}^2$ & $\mathrm{SE}\!\left( \sigma_{r:\mathrm{ad}}^2 \right)$ &
  $\sigma_{L:\mathrm{ad}}^2$ & $\mathrm{SE}\!\left( \sigma_{L:\mathrm{ad}}^2 \right)$ &
  $\sigma_{R:\mathrm{ad}}^2$ & $\mathrm{SE}\!\left( \sigma_{R:\mathrm{ad}}^2 \right)$ \\\hline
& &  & True value 
  & $1.00$ &  & $0.25$ &  & $1.25$ &  &
    &  &  &  &  &  \\\hline
 $3$ & $3$ & $0.25$ & ANOVA
 & $0.989$ & $0.495$ & $0.251$ & $0.475$ & $1.240$ & $0.546$
                             &  &  &  &  &  &  \\
 &   &   & \emph{boot-$i$}
& $0.972$ & $0.364$ & $0.051$ & $0.316$ & $1.023$ & $0.403$
& $1.459$ & $0.547$ & $0.076$ & $0.473$ & $1.535$ & $0.604$ \\
 &   &   & \emph{boot-$j_r$}
& $0.675$ & $0.335$ & $0.611$ & $0.550$ & $1.286$ & $0.503$
& $1.012$ & $0.503$ & $0.274$ & $0.644$ & $1.286$ & $0.503$ \\
 &   &   &\emph{boot-$j_s$}
& $0.645$ & $0.366$ & $0.610$ & $0.549$ & $1.254$ & $0.470$
& $0.967$ & $0.549$ & $0.287$ & $0.671$ & $1.254$ & $0.470$ \\
 &   &   & \emph{boot-$ij_r$}
& $0.666$ & $0.415$ & $0.366$ & $0.561$ & $1.033$ & $0.591$
& $1.499$ & $0.934$ & $0.050$ & $1.005$ & $1.549$ & $0.886$ \\
 &   &   &\emph{boot-$ij_s$}
& $0.659$ & $0.520$ & $0.305$ & $0.616$ & $0.963$ & $0.628$
& $1.482$ & $1.170$ & $-0.037$ & $1.168$ & $1.445$ & $0.943$ \\
\hline
$3$ & $3$ & $2.00$ & ANOVA
& $0.989$ & $0.494$ & $2.014$ & $1.686$ & $3.003$ & $1.724$
&  &  &  &  &  &  \\
 &   &   & \emph{boot-$i$}
& $0.989$ & $0.369$ & $1.216$ & $1.121$ & $2.206$ & $1.165$
& $1.484$ & $0.553$ & $1.825$ & $1.682$ & $3.308$ & $1.747$ \\
 &   &   & \emph{boot-$j_r$}
& $0.653$ & $0.325$ & $2.305$ & $0.908$ & $2.958$ & $0.874$
& $0.980$ & $0.488$ & $1.978$ & $0.974$ & $2.958$ & $0.874$ \\
 &   &   &\emph{boot-$j_s$}
& $0.651$ & $0.371$ & $2.402$ & $0.904$ & $3.053$ & $0.847$
& $0.977$ & $0.556$ & $2.076$ & $0.995$ & $3.053$ & $0.847$ \\
 &   &   & \emph{boot-$ij_r$}
& $0.670$ & $0.424$ & $1.592$ & $1.438$ & $2.262$ & $1.459$
& $1.507$ & $0.954$ & $1.886$ & $2.256$ & $3.393$ & $2.189$ \\
 &   &   &\emph{boot-$ij_s$}
& $0.669$ & $0.524$ & $1.479$ & $1.478$ & $2.149$ & $1.483$
& $1.506$ & $1.179$ & $1.717$ & $2.375$ & $3.223$ & $2.225$ \\
\hline
$50$ & $50$ & $0.25$ & ANOVA
& $1.000$ & $0.029$ & $0.249$ & $0.053$ & $1.249$ & $0.060$
&  &  &  &  &  &  \\
 &     &   & \emph{boot-$i$}
& $1.000$ & $0.028$ & $0.245$ & $0.051$ & $1.245$ & $0.059$
& $1.020$ & $0.029$ & $0.250$ & $0.052$ & $1.270$ & $0.060$ \\
 &     &   & \emph{boot-$j_r$}
& $0.980$ & $0.028$ & $0.271$ & $0.021$ & $1.251$ & $0.034$
& $1.000$ & $0.028$ & $0.251$ & $0.021$ & $1.251$ & $0.034$ \\
 &     &   &\emph{boot-$j_s$}
& $0.979$ & $0.028$ & $0.270$ & $0.021$ & $1.249$ & $0.034$
& $0.999$ & $0.028$ & $0.250$ & $0.021$ & $1.249$ & $0.034$ \\
 &     &   & \emph{boot-$ij_r$}
& $0.980$ & $0.039$ & $0.264$ & $0.055$ & $1.244$ & $0.068$
& $1.020$ & $0.041$ & $0.249$ & $0.056$ & $1.269$ & $0.069$ \\
 &     &   &\emph{boot-$ij_s$}
& $0.980$ & $0.048$ & $0.261$ & $0.059$ & $1.241$ & $0.076$
& $1.020$ & $0.050$ & $0.246$ & $0.060$ & $1.266$ & $0.077$ \\
\hline
$50$ & $50$ & $2.00$ & ANOVA
& $1.001$ & $0.029$ & $1.996$ & $0.399$ & $2.997$ & $0.400$
&  &  &  &  &  &  \\
 &     &   & \emph{boot-$i$}
& $1.000$ & $0.028$ & $1.970$ & $0.387$ & $2.971$ & $0.388$
& $1.021$ & $0.029$ & $2.011$ & $0.395$ & $3.031$ & $0.396$ \\
 &     &   & \emph{boot-$j_r$}
& $0.980$ & $0.028$ & $2.018$ & $0.057$ & $2.998$ & $0.063$
& $1.000$ & $0.028$ & $1.998$ & $0.057$ & $2.998$ & $0.063$ \\
 &     &   &\emph{boot-$j_s$}
& $0.980$ & $0.028$ & $2.013$ & $0.057$ & $2.993$ & $0.063$
& $1.000$ & $0.028$ & $1.993$ & $0.057$ & $2.993$ & $0.063$ \\
 &     &   & \emph{boot-$ij_r$}
& $0.981$ & $0.039$ & $1.966$ & $0.385$ & $2.947$ & $0.387$
& $1.022$ & $0.041$ & $1.986$ & $0.393$ & $3.008$ & $0.395$ \\
 &     &   &\emph{boot-$ij_s$}
& $0.981$ & $0.048$ & $1.983$ & $0.392$ & $2.963$ & $0.396$
& $1.021$ & $0.050$ & $2.003$ & $0.400$ & $3.024$ & $0.404$ \\
\hline
\end{tabular}
\end{table}
\end{landscape}

\begin{table}[htbp]\centering
  \scriptsize
\caption{Lower limits, upper limits, interval ranges, and coverage probabilities of confidence intervals constructed using the standard normal (N), percentile (P), and BCa (B)  methods based on the bootstrap means, for $(n,k)=(5,5)$ and $\sigma_L^2/\sigma_r^2 = 0.05$.}
\label{tab:CI_bootstrap_mean}  
\begin{tabular}{cc|rrrr|rrrr|rrrr}
\hline
 &  & \multicolumn{4}{c|}{$\sigma_{r:(\cdot)}^2$} & \multicolumn{4}{c|}{$\sigma_{L:(\cdot)}^2$} & \multicolumn{4}{c}{$\sigma_{R:(\cdot)}^2$} \\
 Estimator & CI Method & Lower & Upper & Range & \multicolumn{1}{c|}{CP} & Lower & Upper & Range &  \multicolumn{1}{c|}{CP} & Lower & Upper & Range &  \multicolumn{1}{c}{CP} \\
\hline
  Approxi. & ---  & $0.585$ & $2.086$ & $1.500$ & $0.952$ & $-0.028$ & $5.571$ & $5.599$ & $0.952$ & $0.800$ & $3.993$ & $3.193$ & $0.950$ \\
\hline
  \emph{boot-$i$} & N
 & $0.508$ & $1.486$ & $0.978$ & $0.802$ 
 & $-0.246$ & $0.972$ & $1.217$ & $0.676$
 & $0.613$ & $2.107$ & $1.494$ & $0.753$ \\
 \emph{boot-$j_r$} 
 &  & $0.358$ & $1.254$ & $0.895$ & $0.714$
 & $0.031$ & $1.382$ & $1.351$ & $0.848$
 & $0.788$ & $2.237$ & $1.449$ & $0.774$ \\
\emph{boot-$j_s$} 
 &  & $0.317$ & $1.301$ & $0.983$ & $0.761$
 & $-0.001$ & $1.409$ & $1.410$ & $0.859$
 & $0.815$ & $2.211$ & $1.396$ & $0.760$ \\
 \emph{boot-$ij_r$} 
 &  & $0.193$ & $1.388$ & $1.195$ & $0.793$
 & $-0.315$ & $1.426$ & $1.741$ & $0.939$
 & $0.343$ & $2.349$ & $2.006$ & $0.856$ \\
\emph{boot-$ij_s$} 
 &  & $0.053$ & $1.536$ & $1.483$ & $0.874$
 & $-0.462$ & $1.514$ & $1.976$ & $0.951$
 & $0.197$ & $2.443$ & $2.246$ & $0.872$ \\
\hline
  \emph{boot-$i$} & P
 & $0.552$ & $1.504$ & $0.951$ & $0.796$
 & $-0.179$ & $0.951$ & $1.131$ & $0.665$
 & $0.649$ & $2.072$ & $1.423$ & $0.744$ \\
 \emph{boot-$j_r$} 
 &  & $0.373$ & $1.244$ & $0.871$ & $0.707$
 & $0.173$ & $1.493$ & $1.320$ & $0.827$
 & $0.855$ & $2.287$ & $1.432$ & $0.774$ \\
\emph{boot-$j_s$} 
 &  & $0.296$ & $1.219$ & $0.923$ & $0.697$
 & $0.212$ & $1.558$ & $1.345$ & $0.807$
 & $0.877$ & $2.241$ & $1.365$ & $0.755$ \\
 \emph{boot-$ij_r$} 
 &  & $0.294$ & $1.458$ & $1.164$ & $0.821$
 & $-0.099$ & $1.556$ & $1.656$ & $0.971$
 & $0.504$ & $2.439$ & $1.935$ & $0.880$ \\
\emph{boot-$ij_s$} 
 &  & $0.169$ & $1.619$ & $1.450$ & $0.894$
 & $-0.173$ & $1.716$ & $1.889$ & $0.982$
 & $0.385$ & $2.578$ & $2.193$ & $0.891$ \\
\hline
  \emph{boot-$i$} & B
 & $0.569$ & $1.567$ & $0.998$ & $0.806$
 & $-0.182$ & $0.947$ & $1.129$ & $0.666$
 & $0.642$ & $2.075$ & $1.434$ & $0.744$ \\
 \emph{boot-$j_r$} 
 &  & $0.388$ & $1.272$ & $0.884$ & $0.720$
 & $0.244$ & $1.755$ & $1.512$ & $0.797$
 & $0.913$ & $2.404$ & $1.491$ & $0.795$ \\
\emph{boot-$j_s$} 
 &  & $0.269$ & $1.220$ & $0.951$ & $0.694$
 & $0.278$ & $1.980$ & $1.702$ & $0.776$
 & $0.921$ & $2.346$ & $1.424$ & $0.748$ \\
 \emph{boot-$ij_r$} 
 &  & $0.338$ & $1.615$ & $1.277$ & $0.855$
 & $-0.037$ & $1.870$ & $1.906$ & $0.987$
 & $0.576$ & $2.650$ & $2.075$ & $0.909$ \\
\emph{boot-$ij_s$} 
 &  & $0.218$ & $1.848$ & $1.630$ & $0.907$
 & $-0.089$ & $2.312$ & $2.401$ & $0.994$
 & $0.465$ & $2.886$ & $2.421$ & $0.913$ \\
\hline
\end{tabular}
\end{table}

\begin{table}[htbp]\centering
\scriptsize
\caption{Lower limits, upper limits, interval ranges, and coverage probabilities of confidence intervals constructed using the standard normal, percentile, and BCa methods based on the adjusted formula, for $(n,k)=(5,5)$ and $\sigma_L^2/\sigma_r^2 = 0.05$.}
\label{tab:CI_adjusted}  
\begin{tabular}{cc|rrrr|rrrr|rrrr}
\hline
\multicolumn{2}{c|}{} & \multicolumn{4}{c|}{$\sigma_{r:\mathrm{ad}}^2$}
                      & \multicolumn{4}{c|}{$\sigma_{L:\mathrm{ad}}^2$}
                      & \multicolumn{4}{c}{$\sigma_{R:\mathrm{ad}}^2$} \\
Estimator & CI method & Lower & Upper & Range & \multicolumn{1}{c|}{CP}
                      & Lower & Upper & Range & \multicolumn{1}{c|}{CP}
                      & Lower & Upper & Range & \multicolumn{1}{c}{CP} \\
\hline
Approxi. & --- 
& $0.585$ & $2.086$ & $1.500$ & $0.952$
& $-0.028$ & $5.571$ & $5.599$ & $0.952$
& $0.800$ & $3.993$ & $3.193$ & $0.950$ \\
\hline
\emph{boot-$i$}      & N & $0.635$ & $1.857$ & $1.222$ & $0.819$
           & $-0.307$ & $1.215$ & $1.522$ & $0.733$
           & $0.766$  & $2.634$ & $1.868$ & $0.871$ \\
\emph{boot-$j_r$}     &  & $0.448$ & $1.567$ & $1.119$ & $0.882$
           & $-0.206$ & $1.217$ & $1.423$ & $0.814$
           & $0.788$ & $2.237$ & $1.449$ & $0.774$ \\
\emph{boot-$j_s$}   &  & $0.397$ & $1.626$ & $1.229$ & $0.896$
           & $-0.258$ & $1.262$ & $1.520$ & $0.849$
           & $0.815$ & $2.211$ & $1.396$ & $0.760$ \\
\emph{boot-$ij_r$}     &  & $0.302$ & $2.169$ & $1.867$ & $0.976$
           & $-0.692$ & $1.586$ & $2.278$ & $0.930$
           & $0.429$ & $2.936$ & $2.507$ & $0.957$ \\
\emph{boot-$ij_s$}  &  & $0.083$ & $2.400$ & $2.317$ & $0.990$
           & $-0.906$ & $1.725$ & $2.631$ & $0.955$
           & $0.247$ & $3.054$ & $2.808$ & $0.952$ \\
\hline
\emph{boot-$i$}      & P & $0.690$ & $1.879$ & $1.189$ & $0.798$
           & $-0.224$ & $1.189$ & $1.413$ & $0.726$
           & $0.812$ & $2.590$ & $1.778$ & $0.875$ \\
\emph{boot-$j_r$}      &  & $0.466$ & $1.555$ & $1.089$ & $0.876$
           & $-0.064$ & $1.329$ & $1.393$ & $0.847$
           & $0.855$ & $2.287$ & $1.432$ & $0.774$ \\
\emph{boot-$j_s$}   &  & $0.369$ & $1.524$ & $1.154$ & $0.863$
           & $-0.027$ & $1.420$ & $1.447$ & $0.854$
           & $0.877$ & $2.241$ & $1.365$ & $0.755$ \\
\emph{boot-$ij_r$}     &  & $0.460$ & $2.278$ & $1.818$ & $0.970$
           & $-0.472$ & $1.733$ & $2.205$ & $0.961$
           & $0.631$ & $3.049$ & $2.419$ & $0.963$ \\
\emph{boot-$ij_s$}  &  & $0.263$ & $2.529$ & $2.266$ & $0.989$
           & $-0.617$ & $1.960$ & $2.577$ & $0.982$
           & $0.481$ & $3.222$ & $2.742$ & $0.967$ \\
\hline
\emph{boot-$i$}      & B & $0.712$ & $1.959$ & $1.248$ & $0.789$
           & $-0.228$ & $1.184$ & $1.411$ & $0.726$
           & $0.802$ & $2.594$ & $1.792$ & $0.867$ \\
\emph{boot-$j_r$}      &  & $0.485$ & $1.590$ & $1.105$ & $0.881$
           & $0.014$ & $1.599$ & $1.585$ & $0.853$
           & $0.913$ & $2.404$ & $1.491$ & $0.795$ \\
\emph{boot-$j_s$}   &  & $0.337$ & $1.525$ & $1.188$ & $0.861$
           & $0.045$ & $1.875$ & $1.829$ & $0.861$
           & $0.921$ & $2.346$ & $1.424$ & $0.748$ \\
\emph{boot-$ij_r$}     &  & $0.528$ & $2.524$ & $1.996$ & $0.963$
           & $-0.368$ & $2.097$ & $2.465$ & $0.973$
           & $0.719$ & $3.313$ & $2.593$ & $0.961$ \\
\emph{boot-$ij_s$}  &  & $0.340$ & $2.887$ & $2.547$ & $0.991$
           & $-0.463$ & $2.651$ & $3.114$ & $0.986$
           & $0.581$ & $3.608$ & $3.027$ & $0.968$ \\
\hline
\end{tabular}
\end{table}

\begin{landscape}
\begin{table}[htbp]\centering
\small
  \caption{Confidence interval results for adjusted estimators under extreme combinations of $(k,n)$ and $\sigma_L^2/\sigma_r^2$ using the BCa method.}
\label{tab:CI_n_k_ratio_extreme_val}
  \begin{tabular}{cccc|rrrr|rrrr|rrrr}
\hline
 & &   &  & \multicolumn{4}{c|}{$\sigma_{r:\mathrm{ad}}^2$}
           & \multicolumn{4}{c|}{$\sigma_{L:\mathrm{ad}}^2$}
           & \multicolumn{4}{c}{$\sigma_{R:\mathrm{ad}}^2$} \\
$k$ & $n$ & $\sigma^2/L / sigma^2_r$ & Estimator 
     & Lower & Upper & Range & \multicolumn{1}{c|}{CP}
     & Lower & Upper & Range & \multicolumn{1}{c|}{CP}
     & Lower & Upper & Range & \multicolumn{1}{c}{CP} \\
  \hline
 $3$ & $3$ & $0.25$ & Approxi. 
 & $0.411$ & $4.796$ & $4.385$ & $0.950$
 & $-5.290$ & $22.614$ & $27.905$ & $0.940$
 & $0.498$ & $8.491$ & $7.993$ & $0.966$ \\
 &   &   & \emph{boot-$i$} 
 & $0.592$ & $2.450$ & $1.858$ & $0.678$
 & $-0.861$ & $0.655$ & $1.516$ & $0.536$
 & $0.350$ & $2.347$ & $1.996$ & $0.776$ \\
 &  &   & \emph{boot-$j_r$} 
 & $0.115$ & $1.781$ & $1.666$ & $0.775$
 & $-0.500$ & $2.209$ & $2.709$ & $0.904$
 & $0.472$ & $2.452$ & $1.979$ & $0.794$ \\
 &    &   & \emph{boot-$j_s$} 
 & $0.000$ & $1.563$ & $1.563$ & $0.695$
 & $-0.375$ & $2.173$ & $2.548$ & $0.847$
 & $0.540$ & $2.223$ & $1.683$ & $0.664$ \\
 &  &   & \emph{boot-$ij_r$} 
 & $0.194$ & $3.809$ & $3.616$ & $0.943$
 & $-1.523$ & $2.685$ & $4.208$ & $0.989$
 & $0.286$ & $3.718$ & $3.432$ & $0.958$ \\
 &    &   & \emph{boot-$ij_s$} 
 & $0.013$ & $4.473$ & $4.460$ & $0.944$
 & $-1.884$ & $3.027$ & $4.911$ & $0.967$
 & $0.079$ & $3.675$ & $3.596$ & $0.914$ \\\hline
 $3$ & $3$ & $2.00$ & Approxi.
 & $0.411$ & $4.794$ & $4.383$ & $0.954$
 & $-1.966$ & $92.246$ & $94.211$ & $0.952$
 & $0.986$ & $51.101$ & $50.115$ & $0.919$ \\
  &  &  & \emph{boot-$i$} 
 & $0.624$ & $2.504$ & $1.880$ & $0.682$
 & $-0.844$ & $3.841$ & $4.685$ & $0.552$
 & $0.386$ & $5.459$ & $5.073$ & $0.669$ \\
  &  &  & boot-i$'$ 
 & $1.020$ & $6.956$ & $5.936$ & $0.576$
 & $-1.743$ & $3.136$ & $4.879$ & $0.520$
 & $0.954$ & $5.357$ & $4.403$ & $0.661$ \\
  &  &  & \emph{boot-$j_r$} 
 & $0.116$ & $1.724$ & $1.608$ & $0.741$
 & $0.710$ & $4.704$ & $3.995$ & $0.592$
 & $1.465$ & $4.864$ & $3.399$ & $0.510$ \\
  &  &  & \emph{boot-$j_s$} 
 & $0.000$ & $1.581$ & $1.581$ & $0.693$
 & $1.022$ & $4.798$ & $3.775$ & $0.548$
 & $1.724$ & $4.699$ & $2.975$ & $0.439$ \\
  &  &  & \emph{boot-$ij_r$} 
 & $0.195$ & $3.901$ & $3.705$ & $0.938$
 & $-1.278$ & $6.923$ & $8.201$ & $0.884$
 & $0.314$ & $7.882$ & $7.568$ & $0.865$ \\
  &  &  & \emph{boot-$ij_s$}
 & $0.016$ & $4.499$ & $4.484$ & $0.958$
 & $-1.733$ & $7.177$ & $8.911$ & $0.854$
 & $0.096$ & $7.823$ & $7.727$ & $0.840$ \\
 \hline
 $50$ & $50$ & $0.25$ & Approxi.
 & $0.946$ & $1.058$ & $0.112$ & $0.951$
 & $0.168$ & $0.398$ & $0.230$ & $0.950$
 & $1.137$ & $1.379$ & $0.242$ & $0.944$ \\
  &   &   & \emph{boot-$i$} 
 & $0.966$ & $1.079$ & $0.113$ & $0.876$
 & $0.167$ & $0.380$ & $0.213$ & $0.915$
 & $1.169$ & $1.408$ & $0.239$ & $0.911$ \\
  &   &   & \emph{boot-$j_r$}
 & $0.947$ & $1.057$ & $0.110$ & $0.940$
 & $0.215$ & $0.299$ & $0.084$ & $0.559$
 & $1.190$ & $1.325$ & $0.135$ & $0.699$ \\
  &   &   & \emph{boot-$j_s$}
 & $0.943$ & $1.054$ & $0.111$ & $0.946$
 & $0.215$ & $0.302$ & $0.087$ & $0.575$
 & $1.188$ & $1.323$ & $0.135$ & $0.723$ \\
  &   &   & \emph{boot-$ij_r$} 
 & $0.945$ & $1.104$ & $0.159$ & $0.971$
 & $0.160$ & $0.391$ & $0.231$ & $0.943$
 & $1.152$ & $1.428$ & $0.275$ & $0.956$ \\
  &   &   & \emph{boot-$ij_s$} 
 & $0.929$ & $1.125$ & $0.196$ & $0.996$
 & $0.154$ & $0.403$ & $0.249$ & $0.961$
 & $1.137$ & $1.448$ & $0.310$ & $0.980$ \\
 \hline
 $50$ & $50$ & $2.00$ & Approxi. 
 & $0.947$ & $1.059$ & $0.112$ & $0.958$
 & $1.387$ & $3.111$ & $1.724$ & $0.949$
 & $2.334$ & $3.992$ & $1.658$ & $0.943$ \\
  &  &  & \emph{boot-$i$}
 & $0.967$ & $1.079$ & $0.112$ & $0.879$
 & $1.386$ & $2.995$ & $1.609$ & $0.941$
 & $2.404$ & $4.016$ & $1.612$ & $0.936$ \\
  &  &  & \emph{boot-$j_r$} 
 & $0.947$ & $1.057$ & $0.111$ & $0.946$
 & $1.899$ & $2.124$ & $0.224$ & $0.227$
 & $2.888$ & $3.137$ & $0.249$ & $0.243$ \\
  &  &  & \emph{boot-$j_s$}
 & $0.945$ & $1.056$ & $0.111$ & $0.954$
 & $1.896$ & $2.120$ & $0.225$ & $0.196$
 & $2.883$ & $3.131$ & $0.248$ & $0.225$ \\
  &  &  & \emph{boot-$ij_r$} 
 & $0.946$ & $1.106$ & $0.160$ & $0.968$
 & $1.363$ & $2.966$ & $1.603$ & $0.927$
 & $2.380$ & $3.992$ & $1.611$ & $0.925$ \\
  &  &  & \emph{boot-$ij_s$}
 & $0.929$ & $1.126$ & $0.197$ & $0.995$
 & $1.373$ & $3.014$ & $1.641$ & $0.934$
 & $2.388$ & $4.039$ & $1.651$ & $0.935$ \\
\hline
\end{tabular}
\end{table}
\end{landscape}

\begin{table}[htbp]
  \centering
  \caption{Interlaboratory study results for manganese content in iron ore (ISO 5725-4)}
  \label{tab:ex_ISO5725}
  \begin{tabular}{|c|c|c|c|c|}
    \hline
     & \multicolumn{4}{c|}{Replicate} \\
    \cline{2-5}
    Laboratory           & 1 & 2 & 3 & 4 \\
    \hline
     1  & 0.0249 & 0.0259 & 0.0249 & 0.0246 \\
     2  & 0.0316 & 0.0313 & 0.0308 & 0.0315 \\
     3  & 0.0222 & 0.0224 & 0.0271 & 0.0273 \\
     4  & 0.0271 & 0.0290 & 0.0288 & 0.0276 \\
     5  & 0.0271 & 0.0271 & 0.0271 & 0.0271 \\
     6  & 0.0244 & 0.0267 & 0.0251 & 0.0252 \\
     7  & 0.0269 & 0.0283 & 0.0270 & 0.0260 \\
     8  & 0.0272 & 0.0263 & 0.0279 & 0.0265 \\
     9  & 0.0268 & 0.0272 & 0.0274 & 0.0275 \\
    10  & 0.0293 & 0.0304 & 0.0292 & 0.0301 \\
    11  & 0.0311 & 0.0306 & 0.0304 & 0.0294 \\
    12  & 0.0259 & 0.0263 & 0.0250 & 0.0257 \\
    \hline
  \end{tabular}
\end{table}

\begin{table}[htbp]
  \centering
  \caption{Point estimates for the practical example (unit: $10^{-7}\,\%^{2}$).
    From top to bottom: bootstrap mean, bootstrap bias-corrected estimator, and adjusted estimator.
  Values are scaled by $10^{-7}$. 
  ``cor'' denotes the bootstrap bias-corrected estimator; ``ad'' denotes the adjusted estimator based on the bias-adjustment formulas.
  For the bias-corrected block, the SE equals that of the corresponding bootstrap mean and is therefore omitted.}
  \label{tab:ex_point_estimates}
  \small
  \begin{tabular}{|c|rr|rr|rr|}\hline
    & 
    $\hat{\sigma}^{2}_{r:(\cdot)}$ &
    $\widehat{\mathrm{SE}}\!\left(\hat{\sigma}^{2}_{r:(\cdot)}\right)$ &
    $\hat{\sigma}^{2}_{L:(\cdot)}$ &
    $\widehat{\mathrm{SE}}\!\left(\hat{\sigma}^{2}_{L:(\cdot)}\right)$ &
    $\hat{\sigma}^{2}_{R:(\cdot)}$ &
    $\widehat{\mathrm{SE}}\!\left(\hat{\sigma}^{2}_{R:(\cdot)}\right)$ \\\hline
    \textbf{ANOVA}        & 10.77 &  2.47 & 42.73 & 17.83 & 53.51 & 17.91 \\\hline
    \textbf{Bootstrap mean} &
    $\hat{\sigma}^{2}_{r:(\cdot)}$ &
    $\widehat{\mathrm{SE}}\!\left(\hat{\sigma}^{2}_{r:(\cdot)}\right)$ &
    $\hat{\sigma}^{2}_{L:(\cdot)}$ &
    $\widehat{\mathrm{SE}}\!\left(\hat{\sigma}^{2}_{L:(\cdot)}\right)$ &
    $\hat{\sigma}^{2}_{R:(\cdot)}$ &
    $\widehat{\mathrm{SE}}\!\left(\hat{\sigma}^{2}_{R:(\cdot)}\right)$ \\\hline
    \emph{boot-$i$}           & 10.77 &  5.96 & 39.18 & 13.41 & 49.94 & 14.82 \\
    \emph{boot-$j_r$}         &  8.04 &  2.24 & 45.62 &  7.16 & 53.66 &  6.68 \\
    \emph{boot-$j_s$}       &  8.16 &  2.36 & 45.57 &  6.86 & 53.73 &  6.68 \\
    \emph{boot-$ij_r$}         &  8.12 &  5.22 & 40.85 & 15.57 & 48.96 & 16.45 \\
    \emph{boot-$ij_s$}      &  8.06 &  5.69 & 41.83 & 15.93 & 49.89 & 17.18 \\
    \hline
\textbf{Bias-corrected estimator}    &
    $\hat{\sigma}^{2}_{r:\mathrm{cor}}$ &&
    $\hat{\sigma}^{2}_{L:\mathrm{cor}}$ &&
    $\hat{\sigma}^{2}_{R:\mathrm{cor}}$ &\\
    \hline
    \emph{boot-$i$}           & 10.78 && 46.29 && 57.07 & \\
    \emph{boot-$j_r$}         & 13.50 && 39.85 && 53.35 &\\
    \emph{boot-$j_s$}       & 13.38 && 39.90 && 53.28 &\\
    \emph{boot-$ij_r$}         & 13.43 && 44.62 && 58.05 &\\
    \emph{boot-$ij_s$}      & 13.49 && 43.64 && 57.12 &\\
    \hline
    \textbf{Adjusted estimator} &
    $\hat{\sigma}^{2}_{r:\mathrm{ad}}$ &
    $\widehat{\mathrm{SE}}\!\left(\hat{\sigma}^{2}_{r:\mathrm{ad}}\right)$ &
    $\hat{\sigma}^{2}_{L:\mathrm{ad}}$ &
    $\widehat{\mathrm{SE}}\!\left(\hat{\sigma}^{2}_{L:\mathrm{ad}}\right)$ &
    $\hat{\sigma}^{2}_{R:\mathrm{ad}}$ &
    $\widehat{\mathrm{SE}}\!\left(\hat{\sigma}^{2}_{R:\mathrm{ad}}\right)$ \\
    \hline
    \emph{boot-$i$}           & 11.75 &  6.50 & 42.74 & 14.63 & 54.48 & 16.17 \\
    \emph{boot-$j_r$}         & 10.72 &  2.98 & 42.93 &  7.47 & 53.66 &  6.68 \\
    \emph{boot-$j_s$}       & 10.88 &  3.15 & 42.85 &  7.09 & 53.73 &  6.68 \\
    \emph{boot-$ij_r$}         & 11.81 &  7.59 & 41.61 & 17.08 & 53.42 & 17.95 \\
    \emph{boot-$ij_s$}      & 11.73 &  8.28 & 42.70 & 17.41 & 54.43 & 18.74 \\
    \hline
  \end{tabular}
\end{table}

\begin{landscape}
\begin{table}[htbp]
  \centering
  \caption{95\% confidence intervals for adjusted estimators (unit: $10^{-7}\,\%^{2}$).
  Columns report Lower, Upper, and Width for each variance component.}
  \label{tab:adjusted_CIs_practical_swapped}
  \small
  \begin{tabular}{|cc|rrr|rrr|rrr|}
    \hline
    &&
    \multicolumn{3}{c|}{$\hat{\sigma}^{2}_{r:\mathrm{ad}}$} &
    \multicolumn{3}{c|}{$\hat{\sigma}^{2}_{L:\mathrm{ad}}$} &
    \multicolumn{3}{c|}{$\hat{\sigma}^{2}_{R:\mathrm{ad}}$} \\
    Estimator & CI method &  Lower & Upper & Range & Lower & Upper & Range & Lower & Upper & Rnage \\
     \hline
    Approx. & ---
         & $7.13$  & $18.18$ & $11.05$ & $20.05$ & $128.30$ & $108.20$ & $29.25$ & $127.60$ & $98.39$ \\
     \hline
     \emph{boot-$i$}        & N & $-0.99$ & $24.48$ & $25.47$ & $14.07$ & $71.41$ & $57.34$ & $22.80$ & $86.17$ & $63.37$ \\
     \emph{boot-$j_r$}      & N & $4.88$  & $16.57$ & $11.69$ & $28.30$ & $57.57$ & $29.26$ & $40.56$ & $66.75$ & $26.19$ \\
     \emph{boot-$j_s$}    & N & $4.71$  & $17.06$ & $12.35$ & $28.95$ & $56.75$ & $27.79$ & $40.64$ & $66.83$ & $26.19$ \\
     \emph{boot-$ij_r$}      & N & $-3.07$ & $26.68$ & $29.75$ & $8.14$  & $75.08$ & $66.95$ & $18.24$ & $88.59$ & $70.35$ \\
     \emph{boot-$ij_s$}   & N & $-4.50$ & $27.95$ & $32.44$ & $8.59$  & $76.81$ & $68.23$ & $17.70$ & $91.15$ & $73.46$ \\
     \hline
    \emph{boot-$i$}        & P & $3.68$  & $25.86$ & $22.18$ & $11.41$ & $71.33$ & $59.92$ & $22.94$ & $87.34$ & $64.40$ \\
    \emph{boot-$j_r$}      & P & $3.32$  & $14.50$ & $11.18$ & $32.20$ & $62.36$ & $30.16$ & $41.40$ & $66.72$ & $25.32$ \\
    \emph{boot-$j_s$}    & P & $2.04$  & $14.36$ & $12.33$ & $34.17$ & $61.74$ & $27.57$ & $40.83$ & $67.14$ & $26.31$ \\
    \emph{boot-$ij_r$}      & P & $2.81$  & $30.08$ & $27.26$ & $8.84$  & $76.11$ & $67.27$ & $21.34$ & $92.37$ & $71.03$ \\
    \emph{boot-$ij_s$}   & P & $1.60$  & $32.06$ & $30.45$ & $13.25$ & $82.45$ & $69.20$ & $24.25$ & $96.84$ & $72.59$ \\
    \hline
    \emph{boot-$i$}        & B & $4.35$  & $39.41$ & $35.05$ & $13.62$ & $73.70$ & $60.08$ & $24.86$ & $89.45$ & $64.59$ \\
    \emph{boot-$j_r$}      & B & $3.12$  & $14.29$ & $11.17$ & $34.94$ & $67.35$ & $32.42$ & $41.97$ & $67.28$ & $25.31$ \\
    \emph{boot-$j_s$}    & B & $2.04$  & $14.36$ & $12.33$ & $34.57$ & $71.61$ & $37.04$ & $41.58$ & $69.54$ & $27.96$ \\
    \emph{boot-$ij_r$}      & B & $3.66$  & $41.42$ & $37.76$ & $13.21$ & $83.18$ & $69.97$ & $26.74$ & $97.26$ & $70.53$ \\
    \emph{boot-$ij_s$}   & B & $2.50$  & $38.01$ & $35.51$ & $16.53$ & $91.82$ & $75.29$ & $28.26$ & $108.80$ & $80.56$ \\
    \hline
  \end{tabular}
\end{table}
\end{landscape}

\appendix

\section{Derivation of the adjustment formulas for bootstrap schemes}
\label{sec:deriv-adjustm-form-f}

\end{document}